\documentclass[twocolumn,9pt]{article}
\usepackage{extsizes}
\usepackage[super,sort&compress,comma]{natbib} 
\usepackage[version=3]{mhchem}
\usepackage[left=1.5cm, right=1.5cm, top=1.785cm, bottom=2.0cm]{geometry}
\usepackage{balance}
\usepackage{times}
\usepackage{sectsty}
\usepackage{graphicx} 
\usepackage{lastpage}
\usepackage[format=plain,justification=justified,singlelinecheck=false,font={stretch=1.125,small,sf},labelfont=bf,labelsep=space]{caption}
\usepackage{float}
\usepackage{fancyhdr}
\usepackage{fnpos}
\usepackage[english]{babel}
\usepackage{array}
\usepackage{droidsans}
\usepackage{charter}
\usepackage[T1]{fontenc}
\usepackage[usenames,dvipsnames]{xcolor}
\usepackage{setspace}
\usepackage[compact]{titlesec}
\usepackage{lineno}
\definecolor{cream}{RGB}{222,217,201}

\usepackage{bm}
\usepackage{amssymb}
\usepackage[free-standing-units]{siunitx}
\newcommand{\secref}[1]{Section~\ref{#1}}

\newcommand{\figref}[1]{Fig.~\ref{#1}}

\newcommand{\DLTemp}[0]{\tilde{T}}

\begin{document}

\pagestyle{fancy}
\thispagestyle{plain}
%\fancypagestyle{plain}{

%%%HEADER%%%
%\fancyhead[C]{\includegraphics[width=18.5cm]{head_foot/header_bar}}
%\fancyhead[L]{\hspace{0cm}\vspace{1.5cm}\includegraphics[height=30pt]{head_foot/journal_name}}
%\fancyhead[R]{\hspace{0cm}\vspace{1.7cm}\includegraphics[height=55pt]{head_foot/RSC_LOGO_CMYK}}
%\renewcommand{\headrulewidth}{0pt}
%}
%%%END OF HEADER%%%

%%%PAGE SETUP - Please do not change any commands within this section%%%
\makeFNbottom
\makeatletter
\renewcommand\LARGE{\@setfontsize\LARGE{15pt}{17}}
\renewcommand\Large{\@setfontsize\Large{12pt}{14}}
\renewcommand\large{\@setfontsize\large{10pt}{12}}
\renewcommand\footnotesize{\@setfontsize\footnotesize{7pt}{10}}
\makeatother

\renewcommand{\thefootnote}{\fnsymbol{footnote}}
\renewcommand\footnoterule{\vspace*{1pt}% 
\color{cream}\hrule width 3.5in height 0.4pt \color{black}\vspace*{5pt}} 
\setcounter{secnumdepth}{5}

\makeatletter 
\renewcommand\@biblabel[1]{#1}            
\renewcommand\@makefntext[1]% 
{\noindent\makebox[0pt][r]{\@thefnmark\,}#1}
\makeatother 
\renewcommand{\figurename}{\small{Fig.}}
\sectionfont{\sffamily\Large}
\subsectionfont{\normalsize}
\subsubsectionfont{\bf}
\setstretch{1.125} %In particular, please do not alter this line.
\setlength{\skip\footins}{0.8cm}
\setlength{\footnotesep}{0.25cm}
\setlength{\jot}{10pt}
\titlespacing*{\section}{0pt}{4pt}{4pt}
\titlespacing*{\subsection}{0pt}{15pt}{1pt}
%%%END OF PAGE SETUP%%%

%%%FOOTER%%%
\fancyfoot{}
%\fancyfoot[LO,RE]{\vspace{-7.1pt}\includegraphics[height=9pt]{head_foot/LF}}
%\fancyfoot[CO]{\vspace{-7.1pt}\hspace{13.2cm}\includegraphics{head_foot/RF}}
%\fancyfoot[CE]{\vspace{-7.2pt}\hspace{-14.2cm}\includegraphics{head_foot/RF}}
\fancyfoot[RO]{\footnotesize{\sffamily{1--\pageref{LastPage} ~\textbar  \hspace{2pt}\thepage}}}
\fancyfoot[LE]{\footnotesize{\sffamily{\thepage~\textbar\hspace{3.45cm} 1--\pageref{LastPage}}}}
\fancyhead{}
\renewcommand{\headrulewidth}{0pt} 
\renewcommand{\footrulewidth}{0pt}
\setlength{\arrayrulewidth}{1pt}
\setlength{\columnsep}{6.5mm}
\setlength\bibsep{1pt}
%%%END OF FOOTER%%%

%%%FIGURE SETUP - please do not change any commands within this section%%%
\makeatletter 
\newlength{\figrulesep} 
\setlength{\figrulesep}{0.5\textfloatsep} 

\newcommand{\topfigrule}{\vspace*{-1pt}% 
\noindent{\color{cream}\rule[-\figrulesep]{\columnwidth}{1.5pt}} }

\newcommand{\botfigrule}{\vspace*{-2pt}% 
\noindent{\color{cream}\rule[\figrulesep]{\columnwidth}{1.5pt}} }

\newcommand{\dblfigrule}{\vspace*{-1pt}% 
\noindent{\color{cream}\rule[-\figrulesep]{\textwidth}{1.5pt}} }

\makeatother
%%%END OF FIGURE SETUP%%%

%%%TITLE, AUTHORS AND ABSTRACT%%%
\twocolumn[
  \begin{@twocolumnfalse}
\sffamily
\begin{tabular}{m{18cm} p{0cm} }
\noindent\LARGE{\textbf{Crystallization kinetics of binary colloidal monolayers}}$^\dag$ 
\\%Article title goes here instead of the text "This is the title"
\vspace{0.3cm} & \vspace{0.3cm} \\
\noindent\large{An T. Pham,\textit{$^{a,b}$}$^\ddag$ Ryohei Seto,\textit{$^{c}$}$^\ddag$ Johannes Sch\"onke,\textit{$^{c}$} Daniel Y. Joh,\textit{$^{d}$} Ashutosh Chilkoti,\textit{$^{b,d}$} Eliot Fried$^\ast$\textit{$^{c}$} and Benjamin B. Yellen$^\ast$\textit{$^{a,b,d}$}}\\ %Author names go here instead of "Full name", etc.
 \vspace{0.5cm}
\normalfont
 %#!latexmk main.tex
%
Experiments and simulations are used to study the kinetics of crystal growth 
in a mixture of magnetic and nonmagnetic particles suspended in ferrofluid.
The growth process is quantified using both a bond order parameter 
and a mean domain size parameter.
The largest single crystals obtained in experiments
consist of approximately 1000 particles and form
if the area fraction is held between 
$\SIrange{65}{70}{\percent}$ and the field strength is kept 
in the range of $\SIrange{8.5}{10.5}{Oe}$. 
Simulations indicate that much larger single crystals
containing as many as 5000 particles
can be obtained under impurity-free conditions 
within a few hours. 
If our simulations are modified to include impurity concentrations as small as $\SIrange{1}{2}{\percent}$, then the results agree quantitatively with the experiments. 
These findings provide 
an important step toward developing strategies for growing single crystals 
that are large enough to enable follow-on investigations 
across many subdisciplines in condensed matter physics.
\\
  \vspace{0.5cm}
%} \\%The abstrast goes here instead of the text "The abstract should be..."
\end{tabular}
 \end{@twocolumnfalse} \vspace{0.6cm}
  ]

%%%FONT SETUP - please do not change any commands within this section
\renewcommand*\rmdefault{bch}\normalfont\upshape
\rmfamily
\section*{}
\vspace{-1cm}

%%%FOOTNOTES%%%

\footnotetext{\textit{$^{a}$~Duke University, Department of Mechanical Engineering and Materials Science, Box 90300 Hudson Hall, Durham, NC 27708, USA. E-mail: yellen@duke.edu}}
\footnotetext{\textit{$^{b}$~NSF Research Triangle Material Research Science and Engineering Center (MRSEC), Duke University, Box 90217, Durham, NC 27708, USA}}
\footnotetext{\textit{$^{c}$~Mathematical Soft Matter Unit, Okinawa Institute of Science and Technology Graduate University, Onna, Okinawa, Japan 904-0495. Fax: +81 (0)98 966 1062; Tel: +81 (0)98 966 1318; E-mail: eliot.fried@oist.jp}}
\footnotetext{\textit{$^{d}$~Duke University, Department of Biomedical Engineering, Durham, NC 27708, USA}}
\footnotetext{\dag~Electronic Supplementary Information (ESI) available: Movies of experiment and simulation of rapid quenching. See DOI:10.1039/b000000x/}
\footnotetext{\ddag~These authors contributed equally to this work.}

%%%END OF FOOTNOTES%%%
%\noindent\textit{\small{\textbf{Received Xth XXXXXXXXXX 20XX, Accepted Xth XXXXXXXXX 20XX\newline
%First published on the web Xth XXXXXXXXXX 20XX}}}

%\noindent \textbf{\small{DOI: 10.1039/b000000x}}
%\vspace{0.6cm}

%Footnotes
%Please use \dag to cite the ESI in the main text of the article.
%If you article does not have ESI please remove the the \dag symbol from the title and the above footnotetext.

%%%%%%%%%%%%%%%%%%%%%%%%%%%%%%%%%%%%%%%%%%%%%%%%%%%%%%%
%\input{abstract} %%% When latexdiff is used, we need to put the abstract here.
%\linenumbers
%#!latexmk main.tex
%%%1. Introduction%%%%%%%%%%%%
\section{Introduction}

In the last few decades, there has been great interest in colloidal self-assembly, 
not just for its applications in photonic bandgaps,\cite{Kim_2011,Biswas_1998} 
biosensors,\cite{Pankhurst_2003, Haun_2010}
and templates for advanced manufacturing,\cite{Fredriksson_2007}
but also for its unique ability to mimic diverse phenomena in condensed matter physics.
The experimentally accessible length, time, and energy scales of colloidal particles 
(which are respectively ${\sim}\,\SI{1}{\micro\metre}$, ${\sim}\,\SI{1}{\second}$, and ${\sim}\,k_{\mathrm{B}} T $) 
allow them to be tracked using an optical microscope with single-particle resolution 
and to achieve equilibrium at room temperature.
The ability to tune interparticle interactions makes it possible to observe and probe collective ensemble behavior analogous to processes that occur at the atomic scale. 
For example, colloidal suspensions have been used to study condensed matter phenomena such as spinodal decomposition,\cite{Varrato_2012}
crystal nucleation,\cite{Peng_2015}
point defect dynamics,\cite{Pertsinidis_2001}
grain boundary motion,\cite{Edwards_2014}
and glass transitions,\cite{Pham_2002,Tanaka_2010}
along with various liquid-solid and solid-solid transformations.\cite{Yang_2015,Yethiraj_2003} 

%%%%%%%%%%%%%%%%%%%%%%%%%%%%%%%%%%%%%%%%%%%%%%%%%%%%%%%%%%%%%%%%%%%%%%%%%%%%%%%%%%%%%%%%%%%%%%%%%%%%
The vast majority of prior investigations of colloidal self-assembly 
have focused on mono-component colloidal suspensions, 
in which both short-range and long-range interactions combine to form large, 
close-packed single crystals.\cite{Pieranski_1980,Pieranski_1983,Bosse_2008}
More recently, interest in studying multi-component colloidal crystals has developed.
These systems exhibit a greater diversity of phases and allow for the study of a more diverse spectrum of phase transitions 
and condensed matter phenomena.
For example, recent experimental work on the three-dimensional (3D) self-assembly of binary colloidal suspensions points to rich phase behavior and various types of diffusionless phase transformations.\cite{Leunissen_2005,Shevchenko_2006,Yethiraj_2003,Casey_2012}
Processes involving the self-assembly of multi-component suspensions have also been explored in two-dimensional (2D) settings, which are simpler to study due to the ease of particle tracking and the ability to simulate large-scale phenomena.
Moreover, 2D systems are interesting in their own right since they incorporate physics and scaling laws that differ from those arising in 3D systems. 
For example, the long-range translational and orientational ordering that favors crystal formation in 3D systems is absent in 2D systems,
where it hinders the growth of large crystals due to the dominance of long wavelength fluctuations.\cite{Gasser_2009,Gray_2015}
Improved understanding of 2D crystallization processes will therefore have an impact in various applications, such as protein crystallization processes occurring in lipid membranes,\cite{Richter_2005} the formation of self-assembled monolayers (SAMs),\cite{ballauff2016self, Sharon_2016} and the construction of novel 2D materials.\cite{qiu2016uniform}

%for complementary analysis with large scale simulations.
%%%%%%%%%%%%%%%%%%%%%%%%%%%%%%%%%%%%%%%%%%%%%%%%%%%%%%%%%%%%%%%%%%%%%%%%%%%%%%%%%%%%%%%%%%%%%%%%%%%%

To date, there have been very few examples of large single crystals formed in 2D systems of binary colloidal suspensions.
One example, involving electrostatically repulsive particles,\cite{Law_2011} exhibits long-range hexagonal ordering in the absence of close packing.
The structures that form in this setting resemble those typically observed in mono-component colloidal systems.\cite{Pieranski_1980,Pieranski_1983} 
Binary crystals assembled under long-range magnetic repulsion have previously been studied.\cite{Ebert_2008}
As a consequence of polydispersity, these systems tend to form small crystals that coexist with the glassy phase.
On the other hand, it has been shown that large single crystals are difficult to form in close-packed binary colloidal suspensions, which provide more relevant models for condensed matter physics.
Recently, our group has assembled binary colloidal crystals in a uniform magnetic field;\cite{Yang_2015}
however, the crystals we obtained were limited in size due to our limited understanding of the mechanisms underlying the kinetics of the assembly process and, thus, a lack of viable strategies for controlling the rate of growth.

%%%%%%%%%%%%%%%%%%%%%%%%%%%%%%%%%%%%%%%%%%%%%%%%%%%%%%%%%%%%%%%%%%%%%%%%%%%%%%%%%%%%%%%%%%%%%%%%%%%%

Here we study the kinetics of crystal growth in a binary colloidal suspension of magnetic 
and nonmagnetic particles immersed in ferrofluid. 
In a uniform magnetic field aligned normal to the 2D suspension, 
like particles attract and unlike particles repel in a manner similar 
to particles bearing equal and opposite charges, which favor the formation of a self-assembled checkerboard lattice.
The strength of these interactions can be controlled by the external magnetic field.
Since the effective temperature of this system is inversely proportional to the square of the strength of that field, the annealing conditions can easily be tuned with the external magnetic field.
Due to the absence of shielding or other external effects, experimental system 
can be adequately represented by a model of interacting point dipoles.

%%%%%%%%%%%%%%%%%%%%%%%%%%%%%%%%%%%%%%%%%%%%%%%%%%%%%%%%%%%%%%%%%%%%%%%%%%%%%%%%%%%%%%%%%%%%%%%%%%%%

Our goal here is to carefully analyze the time evolution of crystal growth 
as a function of effective temperature and density,
with the objective of determining the ideal range of conditions to form 
the largest possible single crystals.
Our idealized simulations suggest that it is possible to form single crystals containing in excess of several thousand particles;
however, our experiments rarely achieve such sizes due to the presence of particle impurities in concentrations even as small as ${\sim}\,\SIrange{1}{2}{\percent}$.
When we include similar impurity concentrations in simulations, 
we find excellent agreement with experimental behavior.

%%%%%%%%%%%%%%%%%%%%%%%%%%%%%%%%%%%%%%%%%%%%%%%%%%%%%%%%%%%%%%%%%%%%%%%%%%%%%%%%%%%%%%%%%%%%%%%%%%%%%%%

The remainder of the paper is structured as follows.
\secref{sec:experiment_method} provides concise descriptions of the materials used in our experiments, the experimental setup and measurement techniques, and the methods of data analysis.
Theoretical and simulation models are presented in \secref{sec:simulation_model}.
In \secref{sec:results_discussion}, we compare the results from experiments and simulations, 
including order parameters, crystal sizes, and the impact of impurities on the kinetics of crystal growth. 
Finally, we briefly provide concluding remarks and discuss directions for future work.

\begin{figure}[t]
  \centering
  \includegraphics[width=\columnwidth]{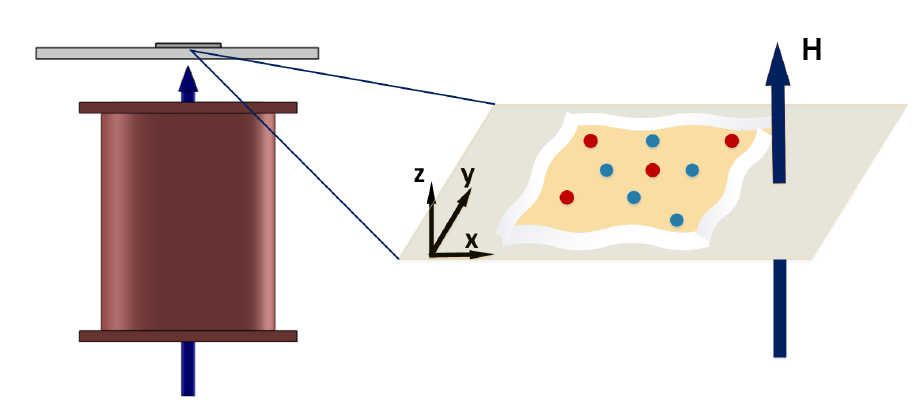}
  \caption{%
    The experimental setup consists of a vertical solenoid with an experimental sample mounted above.
    The magnified view on the right provides an illustration of the experimental sample with magnetic particles shown in blue and nonmagnetic particles shown in red. 
    A magnetic field $\bm{H}$(arrow) is applied in the vertical direction.}
  \label{Experiment Apparatus}
\end{figure}

%#!latexmk main.tex
\section{Experiment methods and data analysis}
\label{sec:experiment_method}

The experimental system involves a binary mixture of magnetic and nonmagnetic spherical particles.
Specifically,
the magnetic particles 
(M-270 Dynabeads\textsuperscript{\textregistered{}}, Life Technologies\texttrademark{})
are of mean diameter $\SI{2.8}{\micro\metre}$ and the nonmagnetic particles (Fluro-Max R0300, Thermo Fischer\texttrademark{})
are of mean diameter $\SI{3.1}{\micro\metre}$.
These particles are mixed with a ferrofluid 
(EMG 705, Ferrotec\texttrademark{}, Bedford, NH) 
which is adjusted so that the volume fraction of magnetic nanoparticles 
is fixed at  ${\sim} \,\SI{1}{\percent}$ in all experiments. 
To reduce nonspecific adhesion of particles to substrates, 
we grow a $\SI{50}{\nano\metre}$ thick poly(oligo(ethylene glycol) methyl ether methacrylate) (POEGMA) polymer brush on 
the glass slides using a surface-initiated, atom-transfer radical polymerization technique (SI-ATRP).\cite{Hucknall_2009} 
Though this polymer brush layer significantly reduces the percentage of particles that 
are randomly immobilized on the substrate, we still occasionally observe 
a few particles stuck to the substrate (typically around ${\sim}\,\SI{0.1}{\percent}$). 
In addition to random binding to the surface, 
we occasionally observe irreversible binding between nonmagnetic particles, 
which represents another type of particle impurity in the experimental system. 
We also observe some rare, giant particle contaminants in the magnetic and nonmagnetic particle 
%suspensions, which are consistently observed in all 
stock concentrations provided by the vendors. 
Due to the compression forces applied when preparing experimental samples, 
these giant particles deform plastically into shapes that resemble oblate spheroids.
These constitute a third type of impurity.
Due to our inability to perfectly control the mixing ratio of the two particle types, 
our experimental samples often have a slight excess of one particle type which impedes the growth 
of single crystals in a manner similar to the effect of the other mentioned impurities.
%%%%%%%%%%%%%%%%%%%%%%%%%%%%%%%%%%%%%%%%%%%%%%%%%%%%%%%%%%%%%%%%%%%%%%%%%%%%%%%%%%%%%%%%%%%%%%%%%%%%

\figref{Experiment Apparatus} provides a schematic of the experimental apparatus. 
A $\SI{2.7}{\micro\liter}$ aliquot of particle mixture is confined between 
the glass surface and a coverslip and sealed using Loctite\textsuperscript{\textregistered{}} marine epoxy. 
Given the dimensions of a typical coverslip, this leads to fluid film of thickness slightly larger than the diameter of a nonmagnetic particle.   
Although we are not able to accurately measure that thickness, we can estimate it by evaluating the center-center distance of a stacked pair of particles, which sometimes forms within the suspension.\cite{Yang_2015, Yang_2015a}  
Based on this measured distance and knowledge of the actual diameter of the particles, we estimated that the film thickness is in the range of $1.1$--$1.3$ magnetic particle diameters (i.e., $\SIrange{3.1}{3.6}{\micro\metre}$).  

The sample is placed on top of a 3D printed platform stage, 
which includes an air-core solenoid that can produce a uniform magnetic 
field throughout the sample. 
Field strength is adjusted by applying current to the solenoid, 
and is controlled with Labview (National Instruments\texttrademark{}, Version 2014, Austin, Texas). 
An inverted Leica\texttrademark{} DMI6000B microscope (LEICA, Bannockburn, IL), 
is used to image the self-assembly process with a $40{\times}$ objective. 
The microscope is capable of automated focusing, and images are captured 
at the rate of two frames per minute with a Qimaging Micropublisher\texttrademark{} 5.0 RTV Camera with resolution 
of $2560{\times}1920$ pixels (Qimaging, Surrey, Canada).

\begin{figure}[t]
  \centering
  \includegraphics[width=\columnwidth]{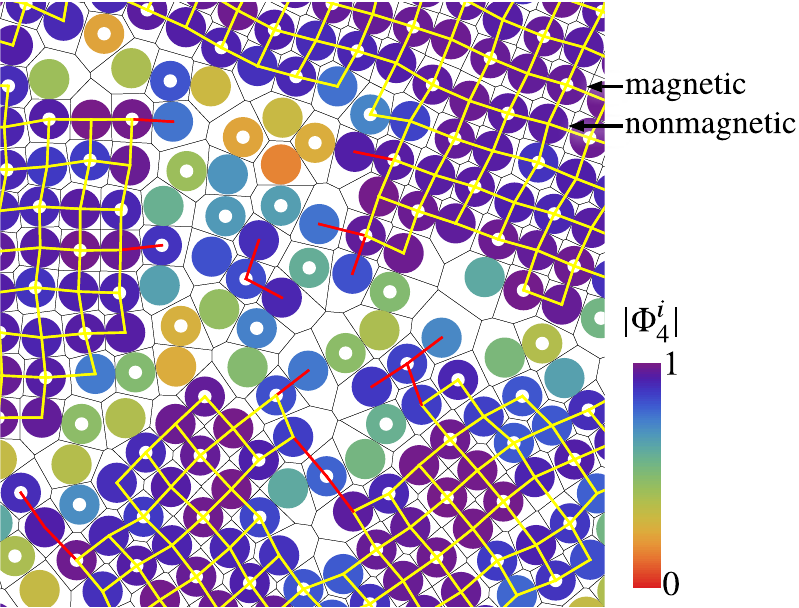}
  \caption{%
    Particles are colored according to the absolute value of 
    the bond order parameter $|\Phi_4^{i}|$.
    Disks with and without holes represent the magnetic and nonmagnetic particles, respectively.
    Correlation bonds are shown in yellow or red. 
    Bonds that form crystal domains
    and that satisfy the criteria described in the text are shown in yellow.
  }
  \label{cd_conditions}
\end{figure}

%%%%%%%%%%%%%%%%%%%%%%%%%%%%%%%%%%%%%%%%%%%%%%%%%%%%%%%%%%%%%%%%%%%%%%%%%%%%%%%%%%%%%%%%%%%%%%%%%%%%%

A custom code is written in MATLAB (Mathworks$\textsuperscript{\copyright}$, Version 2014, Natick, MA) 
for image processing and data analysis.
The circular Hough-transform algorithm\cite{Yuen_1990} 
is used to identify the coordinates of particle centers.
This particle identification algorithm 
is based on the assumption that particles have a spherical shape, 
are of a specific size range, and have a different intensity from that of the background. 
Since the magnetic and nonmagnetic particles are respectively darker and lighter than the ferrofluid,
particle types are distinguished by their average intensity.
Based on particle coordinates extracted in each frame, we then generate Voronoi diagrams 
to ascertain local symmetries,
from which we identify nearest-neighbor bonds for each particle.

%%%%%%%%%%%%%%%%%%%%%%%%%%%%%%%%%%%%%%%%%%%%%%%%%%%%%%%%%%%%%%%%%%%%%%%%%%%%%%%%%%%%%%%%%%%%%%%%%%%%%%%

We use two quantitative measures to characterize the time dependence of the crystallization process, including the average bond order parameter and the mean crystal domain size.
Only particles with nearest neighbor bonds $n_{\mathrm{n}}$ greater than three are included in the bond-order analysis.
We choose this cutoff because it accounts for particles at the edge of a growing crystallite 
but excludes particles at the corners or particles not associated with the crystal. 
The 4-fold bond order parameter of the $i$-th particle is given by
\begin{equation}
  \Phi_{4}^{i} \equiv
  \begin{cases}
    0, & n_\mathrm{n} < 3, \\
    \displaystyle
    \frac{1}{n_\mathrm{n}} \sum_{j=1}^{n_\mathrm{n}} \exp(4 \mathrm{i} \theta_{ji}),
    & n_\mathrm{n} \geq 3,
  \end{cases}
\end{equation}
where $\theta_{ji}$ is the angle between the reference axis and the vector directed from the particle $i$ to the particle $j$, and $\mathrm{i}$ is imaginary number.
Since we observe checkerboard lattices, we use an order parameter with 4-fold symmetry instead of the 6-fold symmetry germane to hexagonally packed crystals.\cite{Strandburg_1988}  
The absolute value of the bond order parameter of a generic particle $i$ in the field of view is averaged according to
\begin{equation}
  \langle \Phi_{4} \rangle \equiv  \frac{1}{N} \sum_{i=1}^{N} | \Phi_{\mathrm{4}}^{i} |,
\end{equation}
which quantifies the degree of 4-fold symmetry in the entire system of $N$ particles.
Large regular crystals generate higher order parameter values, 
with $ \langle \Phi_{4} \rangle = 1$ representing 
an ideal single crystal.

%%%%%%%%%%%%%%%%%%%%%%%%%%%%%%%%%%%%%%%%%%%%%%%%%%%%%%%%%%%%%%%%%%%%%%%%%%%%%%%%%%%%%%%%%%%%%%%%%%%%%%%%%%%%%%%%

In addition to the bond-order analysis, we measure the mean domain size of growing crystals 
based on a modified cluster aggregation parameter originally developed by Vicsek and Family.\cite{vicsek1984dynamic}
The algorithm first evaluates the real part of the bond order parameter 
of the nearest unlike particles, $i$ and $j$.
%$\operatorname{Re}\bigl(\bigr)$
If, for given $i$ and $j$, the real part of the product $\Phi_{4}^{i} \bar\Phi_{4}^{j}$
exceeds a threshold value ($0.6$ is chosen in this work)
and the particle shares the same two correlating neighbors with another particle, 
then all four particles are considered associated with the crystal domain. 
This process is repeated so that each particle is assigned to 
a specific crystal for the entire field of view.
An example resulting from the crystal identification algorithm is shown in \figref{cd_conditions}, where particles connected by yellow line segments are included in the crystal but those connected by red line segments are excluded.
%
%\NEW{We occasionally observed crystallographic defects, such as the point dislocations shown in \figref{cd_conditions}b; however, we do not analyze the dynamical properties of these defects in the present work.}

%%%%%%%%%%%%%%%%%%%%%%%%%%%%%%%%%%%%%%%%%%%%%%%%%%%%%%%%%%%%%%%%%%%%%%%%%%%%%%%%%%%%%%%%%%%%%%%%%%%%%%%%%%%%%%%%%%%%%

We find that the criterion described above robustly identifies 
polycrystalline domains similar to those identified by 
visual inspection of the sample when viewed through the eyepiece of the microscope.
Based on the number of particles included in each domain, 
we then evaluate the mean domain size parameter
\begin{equation}
  \bigl\langle S \bigr\rangle \equiv
  \frac{\sum\limits_{s} s^2 n(s)}{\sum\limits_{s} s n(s)}, 
\end{equation}
where $s$ is the size of domain and $n(s)$ is the number of domains having size $s$.
This calculation averages out size fluctuations between different crystals, 
allowing the determination of a mean domain size within each field of view.

%%%%%%%%%%%%%%%%%%%%%%%%%%%%%%%%%%%%%%%%%%%%%%%%%%%%%%%%%%%%%%%%%%%%%%%%%%%%%%%%%%%%%%%%%%%%%%%%%%%%%%%%%%%%%%%%%%%%%%

The correlation function
\begin{equation}
  g(r) =
  \frac{1}{2 \pi r \Delta{r} \rho (N-1)} \sum_{i \neq j}
  \delta(r - \bigl| \vec{r}_{ij} \bigr|),\label{115529_8Apr16}
\end{equation}
which represents the ratio of the ensemble average of particles within the region between $r$ and $r+\Delta r$ 
and the average number density of the system $\rho = N/L_{x}L_{y}$
can be used to ascertain whether some kinds of impurities are present.
Here, $N$ is the number of particles in the image size of $L_{x} \times L_{y}$.
In a perfect single crystal, the centers of like particles are separated by a distance of $\sqrt{2}$ particle diameters, 
which is the diagonal length in a square lattice. 
Accordingly, peaks of $g(r)$ at smaller separation distances (i.e., $1.0$--$1.1$ particle diameters) between like particles 
indicate the presence of aggregated particles.

%#!latexmk main.tex
\section{Simulation model}
\label{sec:simulation_model}

\subsection{Colloid-colloid interactions}
\label{142724_2Jan16}

A colloidal particle suspension confined in a thin fluid film, having a thickness slightly greater than the typical particle diameter, can be considered as quasi-2D since particle translation is limited to in-plane motion. 
In our experiments, colloidal particles are suspended in a fluid containing iron oxide nanoparticles, called ferrofluid, which we model as a uniform continuum with magnetic permeability $\mu_{\mathrm{f}}$.
%characterized as a function of 
%
We assume that $\mu_{\mathrm{f}}$ is closely approximated in terms of the bulk magnetic susceptibility $\chi_{\mathrm{b}}$ and volume fraction $\varphi$ of the nanoparticles by
\begin{equation}
  \mu_{\mathrm{f}} =
  \mu_{\mathrm{0}}(1 + \varphi \chi_{\mathrm{b}}),
  \label{ferrofluid permeability}
\end{equation}
where $\mu_0$ is the vacuum permeability.
The magnetic moment $\bm{m}_i$ of a spherical particle $i$ 
exposed to a weakly inhomogeneous external magnetic field $\bm{H}$
can be approximated as a magnetic point dipole
\begin{equation}
  \bm{m}_i \approx \bar{\chi}_i \nu_i \bm{H}, 
  \label{induced_magnetic_moment}
\end{equation}
where $\bar{\chi}_i$ is the shape-corrected effective magnetic 
susceptibility of the particle measured relative to that of the surrounding ferrofluid, 
given by
\begin{equation}
  \bar\chi_{i} =
  3\left(\frac{\mu_{i}-\mu_{\mathrm{f}}}
  {\mu_{i}+2\mu_{\mathrm{f}}}
  \right),
  \label{particle susceptibility}
\end{equation}
and $\nu_i$ is the volume of the particle. 
Thus, a magnetic particle effectively behaves as 
paramagnetic when its magnetic permeability is larger than that of the ferrofluid 
and has an effective magnetic susceptibility $\bar{\chi}_{i} > 0$.
A nonmagnetic particle effectively behaves as diamagnetic, 
with susceptibility $\bar{\chi}_{i} < 0$, 
since its magnetic permeability is smaller than that of the ferrofluid. 
The binary suspensions of interest in this work involve two kinds of particles: magnetic particles, with $\bar{\chi}_{\mathrm{m}}$ and $\nu_{\mathrm{m}}$,
and nonmagnetic particles, with $\bar{\chi}_{\mathrm{n}}$ and $\nu_{\mathrm{n}}$.
The volume fraction $\varphi$ of the nanoparticles is controlled to ensure that the effective magnetic moments of the two particle types are equal and opposite, namely to ensure that $\bar{\chi}_{\mathrm{m}} \nu_{\mathrm{m}}  = - \bar{\chi}_{\mathrm{n}} \nu_{\mathrm{n}}$ is satisfied.
In our prior analyses, we have shown that self-consistent magnetic moment calculations lead only to small corrections and can be ignored for computational efficiency.\cite{Yang_2015}
Consistent with these studies, we do not account for these effects here. 
%%%%%%%%%%%%%%%%%%%%%%%%%%%%%%%%%%%%%%%%%%%%%%%%%%%%%%%%%%%%%%%%%%%%%%%%%%%%%%%%%%%%%%%%%%%%%%%%%%%%

\begin{figure}[t]
\centering
 \includegraphics[width=8cm]{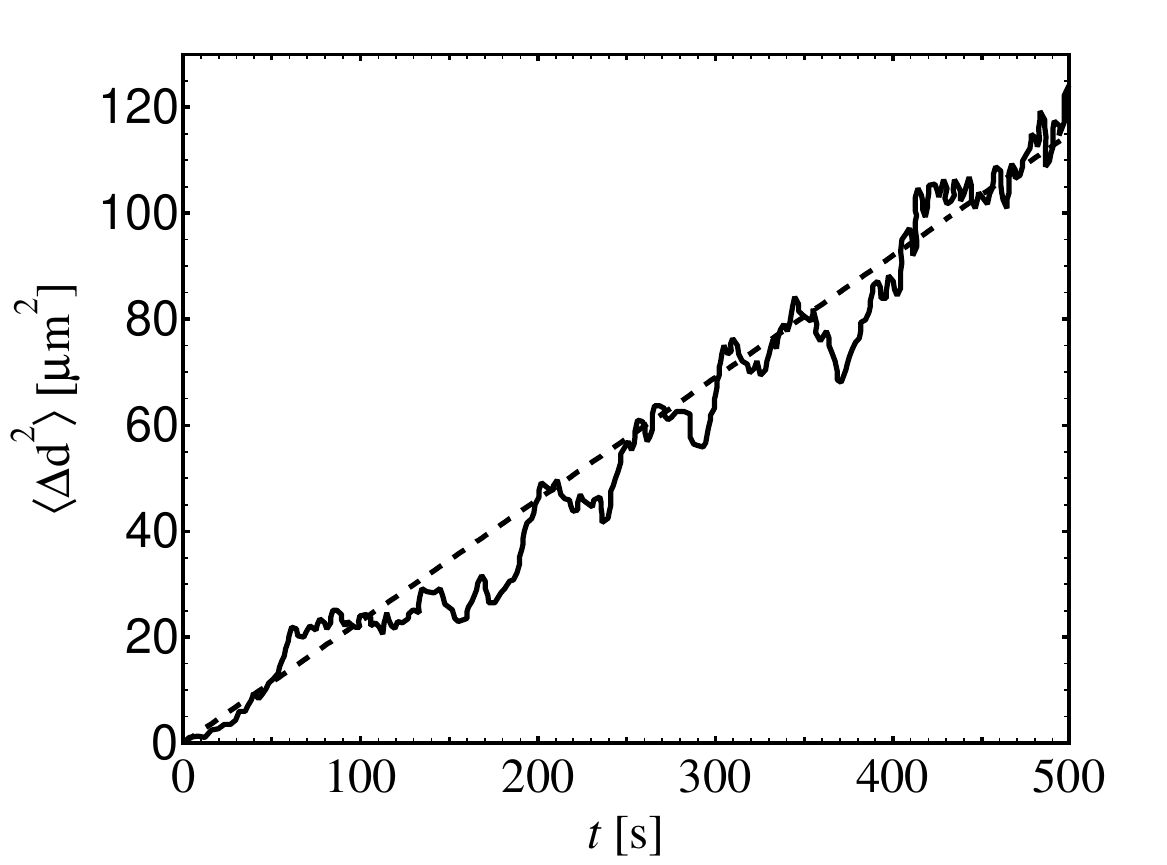}
\caption{%
  The mean square displacement (solid line) averaged for 10 particles in the absence of an external magnetic field is shown as a function of time.
  The diffusion coefficient $D_0 \approx \SI{0.0565}{\square\micro\metre \per \second}$ 
  is obtained through a best fit to experimental data by 
  $\big \langle \Delta{d}^2 \big \rangle = 4 D_0 t \approx 0.226t$.
}
\label{183020_10Dec15}
\end{figure}

The dipole-dipole interaction energy between two colloidal particles $i$ and $j$,
at positions $\bm{X}_i$ and $\bm{X}_j$, is written as
\begin{equation}
  U_{\mathrm{M}}^{(ij)} = 
  -\frac{\mu_{\mathrm{f}}}{4\pi r_{ij}^3} 
\left( 3 (\bm{m}_i \cdot \bm{n}_{ij})
(\bm{m}_j \cdot \bm{n}_{ij}) - \bm{m}_i \cdot \bm{m}_j \right),
  \label{dipole-dipole-potential}
\end{equation}
where $\bm{n}_{ij} \equiv (\bm{X}_j-\bm{X}_i)/r_{ij}$ and $r_{ij} \equiv |\bm{X}_j-\bm{X}_i|$ respectively represent the unit vector and distance between the centers of particles $i$ and $j$.
Since the external field $\bm{H}$ is applied along the $z$-direction 
(i.e., perpendicular to the monolayer),
\eqref{dipole-dipole-potential} is a function only of $r_{ij}$.
This leads to the following magnetic force:
\begin{equation}
 \bm{F}_{\mathrm{M}}^{ij}
 = - \frac{3\mu_{\mathrm{f}}}{4\pi r_{ij}^4}
\bar{\chi}_i \bar{\chi}_j \nu_i  \nu_j H^2 \bm{n}_{ij}. \label{101213_28Jan16}
\end{equation}
Thus, 
whereas the magnetic force between like particles ($\bar{\chi}_{i} \bar{\chi}_{j}  > 0$) is repulsive,
the magnetic force between unlike particles ($\bar{\chi}_{i} \bar{\chi}_{j} < 0$) is attractive.
%
%The ferrofluid concentration is controlled to guarantee that 
%the effective magnetic susceptibilities 
%of the two particle types are equal and opposite: 
%$\bar{\chi}_{\mathrm{m}} \nu_{\mathrm{m}} = - \bar{\chi}_{\mathrm{n}} \nu_{\mathrm{n}}$.

%%%%%%%%%%%%%%%%%%%%%%%%%%%%%%%%%%%%%%%%%%%%%%%%%%%%%%%%%%%%%%%%%%%%%%%%%%%%%%%%%%%%%%%%%%%%%%%%%%%%

Since the colloidal particles are nearly rigid spherical balls, steric effects prevent them from physically overlapping.
To simulate this behavior, 
we employ a soft-sphere model with sufficiently large stiffness $k$, 
in which the following contact force $\bm{F}_{\mathrm{C}}^{(ij)}$ arises 
between overlapping two particles of radii $a_i$ and $a_j$:
\begin{equation}
 \bm{F}_{\mathrm{C}}^{ij} =
\begin{cases}
  0, & r_{ij} > a_i + a_j, \\
 k (r_{ij} - a_i - a_j) \bm{n}_{ij},  & r_{ij} \leq a_i + a_j.
\end{cases}\label{160927_4Apr16}
\end{equation}
%%%%%%%%%%%%%%%%%%%%%%%%%%%%%%%%%%%%%%%%%%%%%%%%%%%%%%%%%%%%%%%%%%%%%%%%%%%%%%%%%%%%%%%%%%%%%%%%%%%%

The colloid-colloid interactions encoded in \eqref{101213_28Jan16} and \eqref{160927_4Apr16} govern the basic features of the system, the net interaction energy of which is minimized when particles form a checkerboard lattice.\cite{Yang_2015} 
However, to investigate the kinetics of crystallization, it is also necessary to account for the solvent's influence on particle dynamics; that influence is discussed next.

\subsection{Solvent-colloid interactions}
\label{143028_5Apr16}

The suspended colloidal particles also interact with the solvent molecules of the host fluid, 
and we employ a Brownian dynamics approach to capture the dominant influences of the solvent.
Since solvent molecules are many orders of magnitude less massive than colloidal particles, it is reasonable to assume that they equilibrate instantaneously. 
Accordingly, we can model their influence with forces that induce viscous dissipation (i.e., hydrodynamic forces) and thermal fluctuations (i.e., Brownian forces).
With this approach, solvent molecules are not explicitly tracked for the sake of computational efficiency.
%

%%%%%%%%%%%%%%%%%%%%%%%%%%%%%%%%%%%%%%%%%%%%%%%%%%%%%%%%%%%%%%%%%%%%%%%%%%%%%%%%%%%%%%%%%%%%%%%%%%%%
The hydrodynamic forces acting on a system of $N$ suspended particles can be represented as a $2N$-dimensional vector $\bm{F}_{\mathrm{H}} \equiv (\bm{F}^1_{\mathrm{H}},\dotsc, \bm{F}^N_{\mathrm{H}})$ of $N$ two-dimensional hydrodynamic force vectors.
In terms of the $2N$-dimensional vector $\bm{U} \equiv (\bm{U}^1,\dotsc,\bm{U}^N)$ of two-dimensional particle velocity vectors,
$\bm{F}_{\mathrm{H}}$ has the form 
%$\bm{U} \equiv (\bm{U}^{(1)}, \dotsc, \bm{U}^{(N)})$;
%
\begin{equation}
 \bm{F}_{\mathrm{H}} = -\bm{R} \bm{U},\label{154600_16Dec15}
\end{equation}
where $\bm{R}$ is a $2N{\times}2N$ resistance matrix.
Note that all force and velocity vectors in this section
are $2N$ dimensional, which represent in-plane components of $N$ particles.

%%%%%%%%%%%%%%%%%%%%%%%%%%%%%%%%%%%%%%%%%%%%%%%%%%%%%%%%%%%%%%%%%%%%%%%%%%%%%%%%%%%%%%%%%%%%%%%%%%%%
Since the Brownian forces $\bm{F}_{\mathrm{B}} \equiv (\bm{F}^1_{\mathrm{B}},\dotsc,\bm{F}^N_{\mathrm{B}})$ are random, their ensemble average must vanish: $\langle \bm{F}_{\mathrm{B}} \rangle = \bm{0}$.
To ensure that the average kinetic energy of each colloidal particle
attains $k_{\mathrm{B}}T$,
the correlations must satisfy\cite{Deutch_1971,Doi_2013}
\begin{equation}
\langle \bm{F}_{\mathrm{B}}(0) \bm{F}_{\mathrm{B}}(t) \rangle
 = 2 k_{\mathrm{B}}T \bm{R} \delta (t),
\label{154607_16Dec15}
\end{equation}
where $\delta$ is the delta function.

%%%%%%%%%%%%%%%%%%%%%%%%%%%%%%%%%%%%%%%%%%%%%%%%%%%%%%%%%%%%%%%%%%%%%%%%%%%%%%%%%%%%%%%%%%%%%%%%%%%%
Since $\bm{R}$ is generally a dense matrix, it is advantageous to introduce an approximation.
% for the resistance matrix $\bm{R}$. 
%%
In a dilute suspension, resistance forces acting on particles are oppositely proportional to their own velocities. 
Thus $\bm{R}$ can be approximated by a diagonal matrix $\bm{R}_{\mathrm{sd}}$:
\begin{equation}
\bm{R} \approx \bm{R}_{\mathrm{sd}} = \zeta \bm{I}.
\label{115317_5Apr16}
\end{equation}
Here, due to the small difference in size between the two species of particles, the suspension is assumed to be monodisperse for simplicity; thus, the resistance coefficient $\zeta$ is the same for all particles.
Due to the presence of confining glass boundaries in our system, a better match with experiment requires $\zeta$ to be determined from experimental measurements of the diffusion coefficient $D_0 \equiv k_{\mathrm{B}} T / \zeta$ of isolated particles. 
The average value $\langle D_0 \rangle \approx \SI{0.0565}{\square\micro\metre \per \second}$ is evaluated from experimental observations (\figref{183020_10Dec15}), leading to an effective Stokes' drag coefficient $\zeta$ about 2.5 times larger than that expected for water. 
%

\begin{figure}[t]
\centering
\includegraphics[width=\columnwidth]{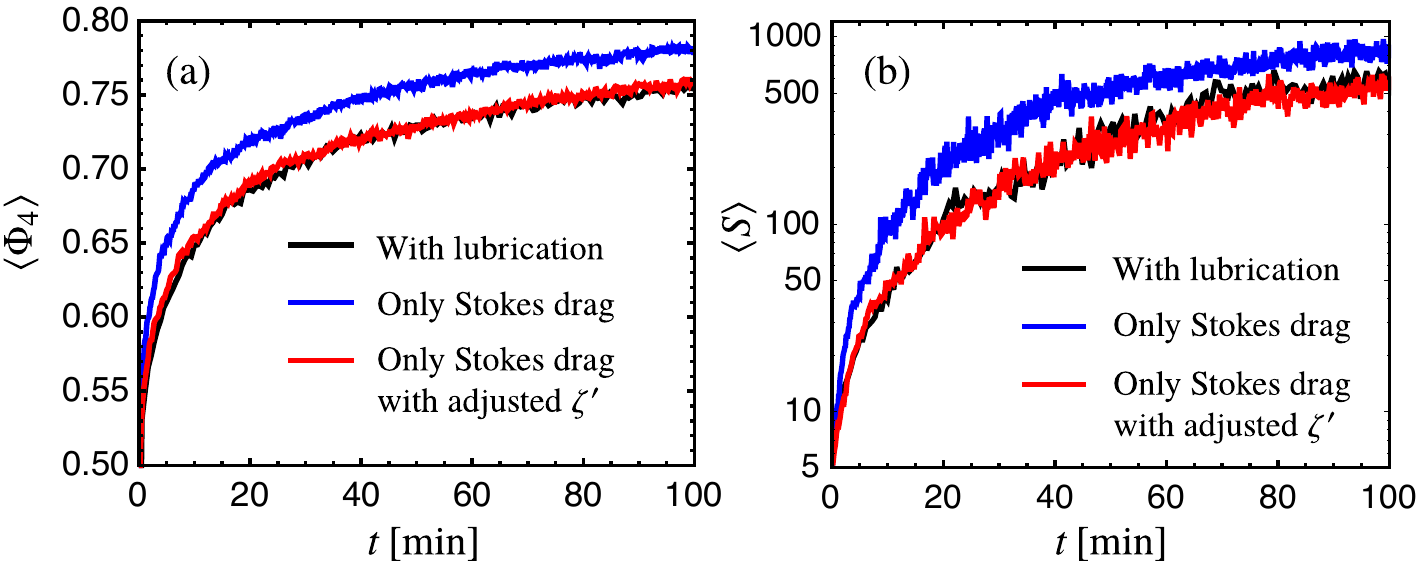} 
\caption{%
  Simulation without lubrication forces (black) compared to a simulation with lubrication forces (blue) using (a) the average order parameter $\langle \Phi_4 \rangle$ and (b) the mean domain size $\langle S \rangle$. 
  The simulation neglecting lubrication forces with an adjusted resistance coefficient $\zeta' \approx 1.9 \zeta$ (red) can capture the features of the simulation with lubrication forces.
}
\label{153003_11Dec15}
\end{figure}

%%%%%%%%%%%%%%%%%%%%%%%%%%%%%%%%%%%%%%%%%%%%%%%%%%%%%%%%%%%%%%%%%%%%%%%%%%%%%%%%%%%%%%%%%%%%%%%%%%%%
In dense suspensions, the hydrodynamic resistance is influenced by the presence of other nearby particles, 
which leads to non-negligible off-diagonal terms in the resistance matrix $\bm{R}$.
Strong hydrodynamic resistances, called lubrication forces, arise from the presence of viscous fluid in the narrow gaps between nearby particles.\cite{Ball_1997} 
%because the resistance coefficient of the normal mode  
%diverges with the inverse of the gap
%
Under these circumstances, $\bm{R}$ may be approximated by an expression that includes the lubrication terms, namely $\bm{R} \approx \bm{R}_{\mathrm{sd}} + \bm{R}_{\mathrm{lub}}$,
where the particular form of $\bm{R}_{\mathrm{lub}}$ can be found elsewhere.\cite{Mari_2014}
If this resistance matrix is used instead of the diagonal alternative \eqref{115317_5Apr16}, 
the computational cost of the simulations is obviously much higher.
Fortunately, when we compare simulations with and without lubrication~(\figref{153003_11Dec15}),
we find that it affects only the time scale of the considered processes.
Therefore, we can rescale the resistance matrix \eqref{115317_5Apr16} in our simulation with an adjusted resistance coefficient $\zeta'$. 
The adjustment factor $ \zeta' /\zeta  \approx 1.9$  is found 
at $\phi_{\mathrm{area}} = 0.68$ for a certain range of field strengths;
specifically, the range of the effective (dimensionless) temperature defined later 
is $0.05 \lesssim \DLTemp \lesssim 0.15$.

\subsection{Equation of motion for colloidal particles}

%We determines the motion of colloidal particles by considering the colloid-colloid
%and colloid-solvent interactions described above.
%

Since particle inertia is neglected, the particles obey a set of quasi-static equations of motion, in which the individual particle trajectories are obtained through a force balance:
\begin{equation}
 \bm{F}_{\mathrm{H}}
+ \bm{F}_{\mathrm{B}}
+ \bm{F}_{\mathrm{M}}
+ \bm{F}_{\mathrm{C}}
= \bm{0},
\label{eq_of_motion}
\end{equation}
%
%These force vectors include all particles like $ \bm{F}_{\mathrm{H}} $,
%defined prior to \eqref{154600_16Dec15}. 
%
where $\bm{F}_{\mathrm{M}} \equiv (\bm{F}^1_{\mathrm{M}},\dots,\bm{F}^N_{\mathrm{M}})$ and $\bm{F}_{\rm C} \equiv (\bm{F}^1_{\rm C},\dots,\bm{F}^N_{\rm C})$ respectively represent the $2N$-dimensional vectors of the magnetic and contact forces. 
At each time step, the magnetic and contact forces, $\bm{F}_{\mathrm{M}}$ and $\bm{F}_{\mathrm{C}}$, and the resistance matrix, $\bm{R}$, are calculated from the particle configuration, and the random Brownian forces, $\bm{F}_{\mathrm{B}}$, satisfying \eqref{154600_16Dec15} are numerically generated. 
Eliminating $\bm{F}_{\mathrm{H}}$ between \eqref{154600_16Dec15} and \eqref{eq_of_motion}, we obtain
\begin{equation}
 \bm{U} = \bm{R}^{-1}
\cdot 
( \bm{F}_{\mathrm{B}} 
+ \bm{F}_{\mathrm{M}} 
+ \bm{F}_{\mathrm{C}}),
\end{equation}
from which we determine the trajectories $\bm{X}$
of the colloidal particles
by integrating with respect to time.\cite{Mari_2015a}

%%%%%%%%%%%%%%%%%%%%%%%%%%%%%%%%%%%%%%%%%%%%%%%%%%%%%%%%%%%%%%%%%%%%%%%%%%%%%%%%%%%%%%%%%%%

The dependence on external field strength $H \equiv |\bm{H}|$ is of essential importance in this work.
On writing $ d = a_{\mathrm{m}} + a_{\mathrm{n}} $, 
the characteristic force scale $F_{\mathrm{M}}^{\ast}$
of the system is given by the magnetic force between magnetic and nonmagnetic particles 
in contact under the field:
\begin{equation}
 F_{\mathrm{M}}^{\ast} \equiv
\frac{3\mu_{\mathrm{f}}\bar{\chi}_{\mathrm{m}} |\bar{\chi}_{\mathrm{n}}| 
\nu_{\mathrm{m}}  \nu_{\mathrm{n}} }{4\pi d^4}  H^2.\label{105111_29Apr16}
\end{equation}
Each term entering the equation of motion~\eqref{eq_of_motion} 
is nondimensionalized based on this force scale: 
for example, $\bm{F}_{\mathrm{H}}$
becomes $\tilde{\bm{F}}_{\mathrm{H}} \equiv \bm{F}_{\mathrm{H}} / F_{\mathrm{M}}^{\ast} $.
In the dynamics, the characteristic time scale 
$\tau_{\mathrm{M}}$ related to the magnetic force scale is given by
\begin{equation} 
 \tau_{\mathrm{M}} \equiv \frac{ \zeta d  }{2 F_{\mathrm{M}}^{\ast}}.\label{160816_8Feb16}
\end{equation}
With force and time scales \eqref{105111_29Apr16} and $\eqref{160816_8Feb16}$, 
the relation \eqref{154607_16Dec15} determining the strength of Brownian forces 
is nondimensionalized in accord with
\begin{equation}
  \bigl\langle \tilde{\bm{F}}_{\mathrm{B}}(0)
\tilde{\bm{F}}_{\mathrm{B}}(\tilde{t}) \bigr\rangle 
%= \frac{2 k_{\mathrm{B}} T}{a F_{\mathrm{M}}^{\ast}}
% \tilde{R} \tilde{\delta}
= 2 \DLTemp \tilde{\bm{R}} \delta(\tilde{t}),
\end{equation}
where $\tilde{t} \equiv  t / \tau_{\mathrm{M}}$ and $\tilde{\bm{R}} \equiv \bm{R}/ \zeta$. 
The dimensionless number 
\begin{equation}
\DLTemp
\equiv 
\frac{2k_{\mathrm{B}}T}{d F_{\mathrm{M}}^{\ast}}
= \frac{8\pi d^3 k_{\mathrm{B}}T}{
3\mu_{\mathrm{f}} \bar{\chi}_{\mathrm{m}} |\bar{\chi}_{\mathrm{n}}|
\nu_{\mathrm{m}}  \nu_{\mathrm{n}} H^2},
\label{eq_dimensionless_temperature}
\end{equation}
therefore serves as an effective temperature.
From \eqref{eq_dimensionless_temperature}, 
we can decrease $\DLTemp$ by increasing the strength $H$ of the external field.
%

%%%% figure*
\begin{figure*}[htb]
  \centering
  \includegraphics[width=0.9\textwidth]{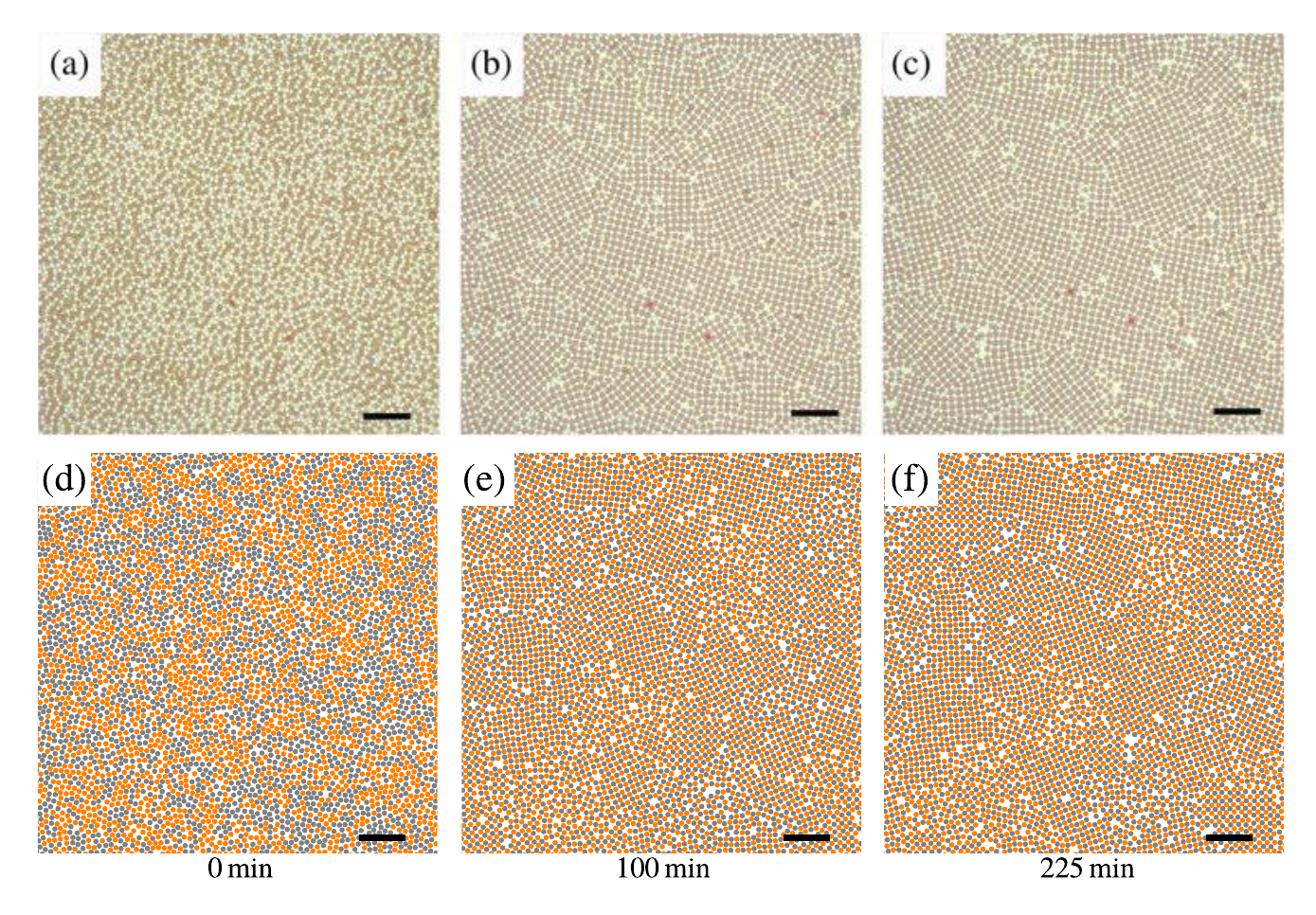}
  \caption{%
    Time evolution of crystal growth during the rapid quenching process at an effective temperature of $\DLTemp \approx 0.081$ ($H=\SI{7.0}{Oe}$) in experiment\,(a--c) and $\DLTemp \approx 0.076$ in simulation\,(d--f). 
    Magnetic (red) and nonmagnetic (white) particles in the experiment\,(a--c) 
    are assigned gray and red colors in the simulation\,(d--f). 
    Images have a size of $\SI{250}{\micro\metre} \times \SI{250}{\micro\metre}$. 
    Scale bars on images show $\SI{25}{\micro\metre}$.
  }
  \label{time_evolution}
\end{figure*}

\subsection{Boundary conditions}
\label{sec_boundary_conditions}

In the experimental system, particles assemble within a fluid film of approximately 
$\SI{20}{\milli\metre}$ (in lateral dimensions). 
The observed field of view is distant enough from any boundaries of the sample
to ensure that boundary effects are practically negligible. 
In simulations, on the other hand, computational limitations require 
that we consider only a small fraction of the system, which
can lead to undesirable boundary effects.
To avoid this potential problem, we use square simulation boxes with periodic boundary conditions.
Additionally, to avoid finite system size effects, which can lead to correlations between a crystal and its periodic images from the system boundaries, we restrict our simulations to systems of greater than $10000$ particles.

\begin{figure*}[htb]
  \centering
  \includegraphics[width=\textwidth]{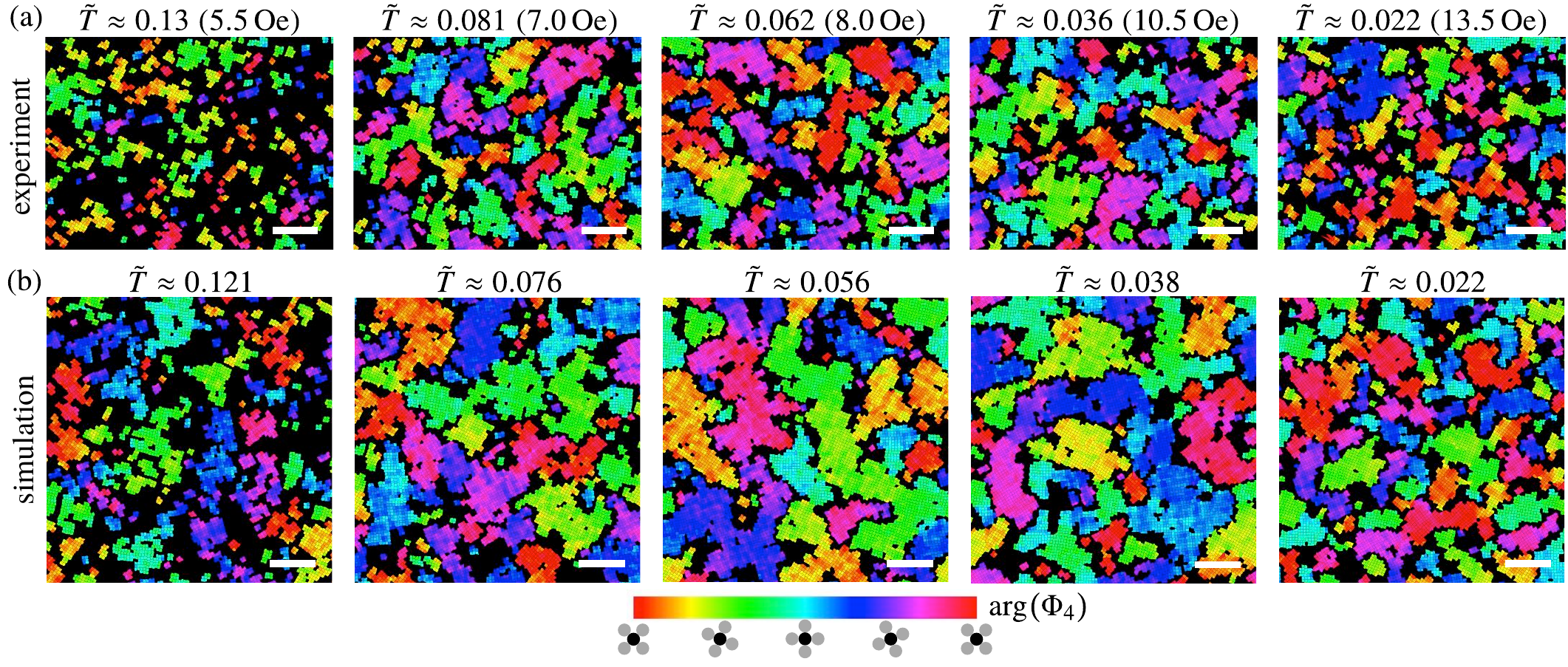}
  \caption{%
    Polycrystalline domains formed in experiments (a) and simulations (b) at different temperatures. 
    Crystal domains are colored according to local orientations of particles. 
    Black coloration depicts the regions where particles do not form crystalline structures.
    The scale bars are $\SI{50}{\micro\metre}$.
  }
  \label{crystal_domain_colored}
\end{figure*}

\section{Results and Discussion}
\label{sec:results_discussion}
\subsection{Order parameter and domain size}
\label{Order parameter and domain size}

To explore the kinetics of crystal growth, we developed an experimental 
system that permits dense monolayers of particles 
to be studied in a static uniform vertical magnetic field. 
\figref{time_evolution} shows typical snapshots of the liquid-solid transition 
obtained from experiments (a--c) and simulations (d--f). 
The experimental field of view is a rectangle with a length and width 
of $85$ and $114$ particle diameters, respectively, 
with each field of view containing approximately $8000$ particles. 
In this figure, the particle concentration is fixed at
$\phi_{\mathrm{area}} = 0.66$, 
while the magnetic field strength $H$ is held constant at $\SI{7}{Oe}$ 
(corresponding to an effective temperature of $\DLTemp \approx 0.081$ in \eqref{eq_dimensionless_temperature}).
In simulations, we model a system commensurate in size to the experimental field of view with $N = 10000$ particles (see \secref{sec_boundary_conditions}). 
Prior to the quenching process, the Brownian simulation with $H = 0$ (i.e., at $\tilde{T} = \infty$) is run for 10 minutes to ensure that the initial configuration is a thermal equilibrium state in which magnetic forces are absent.
Simulation conditions are calibrated to approximate experimental conditions (see Table 1, Movies S1, and S2$^\dag$).

\begin{table}[bht]
  \centering
  \caption{Simulation Parameters}
  \label{tbl:1}
  \begin{tabular}{@{\extracolsep{\fill}}lcr@{\extracolsep{\fill}}}
    \hline 
    %  Parameters & & Values\\
    %  \hline
    magnetic particle diameter  & $d_{\mathrm{m}}$ & $\SI{2.95}{\micro\metre}$ \\
    nonmagnetic particle diameter & $d_{\mathrm{n}}$ & $\SI{2.95}{\micro\metre}$ \\
    magnetic particle's effective susceptibility & $\bar{\chi}_{\mathrm{m}}$  & $0.36$ \\
    nonmagnetic particle's effective susceptibility & $\bar{\chi}_{\mathrm{n}}$  & $-0.36$ \\
    magnetic permeability of ferrofluid  & $\mu_{\mathrm{f}}$ & $1.47\mu_{\mathrm{0}}$ \\
    temperature & $T$       & $\SI{298}{\kelvin}$ \\
    %    magnetic field &$H$              & $4.0$--$\SI{13.5}{Oe}$ & \\
    \hline
  \end{tabular}
\end{table}

%%%%%%%%%%%%%%%%%%%%%%%%%%%%%%%%%%%%%%%%%%%%%%%%%%%%%%%%%%%%%%%%%%%%%%%%%%%%%%%%%%%%%%%%%%%%

The time evolution of crystal growth is shown for a situation in which the system is suddenly exposed to a constant vertical magnetic field at $t = 0$, which mimics a rapid quenching process initiated at infinite temperature.
Applied magnetic fields induce the initial liquid phase to rapidly solidify, nucleating small crystals within the first 20 minutes of experiments. 
In this early stage of the nucleation process, the dominant effect of interfacial energy causes nuclei containing 4--30 particles to spontaneously appear and disappear with time.
Over the next phase of crystal growth, which occurs from $\SIrange{20}{60}{\minute}$, larger crystals survive at the expense of the smaller crystals and the system continues to coarsen until most particles are incorporated into crystals.
At low values of $H$ (close to the solidification temperature), the liquid phase coexists with small crystallites that continue to grow until the end of the experiment. 
At higher values of $H$, the system coarsens until the crystals merge and form domain boundaries, such as those shown in experiment (\figref{time_evolution}c) and simulation (\figref{time_evolution}f).

%%%%%%%%%%%%%%%%%%%%%%%%%%%%%%%%%%%%%%%%%%%%%%%%%%%%%%%%%%%%%%%%%%%%%%%%%%%%%%%%%%%%%%%%%%%%%%%%%%%%%%%%%%%%%%%%%%%%%%%%%%%%%%%%%%

Some examples of polycrystalline structures are provided in \figref{crystal_domain_colored}a (experiments) 
and \figref{crystal_domain_colored}b (simulations), which depict typical domain structures after $225$ minutes of elapsed time for different values of the effective temperature $\DLTemp$.
For ease of visualization, the domains are presented as color coded Voronoi cells to highlight different crystal orientations (i.e., $\arg \Phi_4^{i}$). 
Disordered regions are indicated with black.
The clear dependence of mean crystal size on $\DLTemp$, or, equivalently, the applied magnetic field strength $H$, is evident in both experiments and simulations. 
In a weak magnetic field of $H=\SI{5.5}{Oe}$, corresponding to the solidification temperature $\DLTemp \approx 0.13$, small crystal domains containing 4--100 particles are surrounded by the disordered phase. 
On the other hand, in a strong magnetic field of $H=\SI{13.5}{Oe}$ ($\DLTemp \approx 0.022$), the mean crystal size is equally small but are caused instead by the slow annealing kinetics, which impede crystal growth during the experimental time interval.
Maximum crystal sizes are observed in the medium range of 
$H=\SIrange{8.5}{10.5}{Oe}$ ($\DLTemp \approx 0.036$--$0.055$), 
which is slightly below the solidification temperature. 
This strong dependence of crystal size on quenching conditions indicates the existence of an ideal field strength $H$ for promoting the growth of large single crystals. 
%n
The value of $H$ should be high enough to maintain strong particle association 
with the crystal but weak enough to allow defects to 
diffuse out of the crystal interior, as well as to allow domain boundaries to merge and reform.

%%%% figure*
\begin{figure*}[htb]
  \centering
  \includegraphics[width=0.8\textwidth]{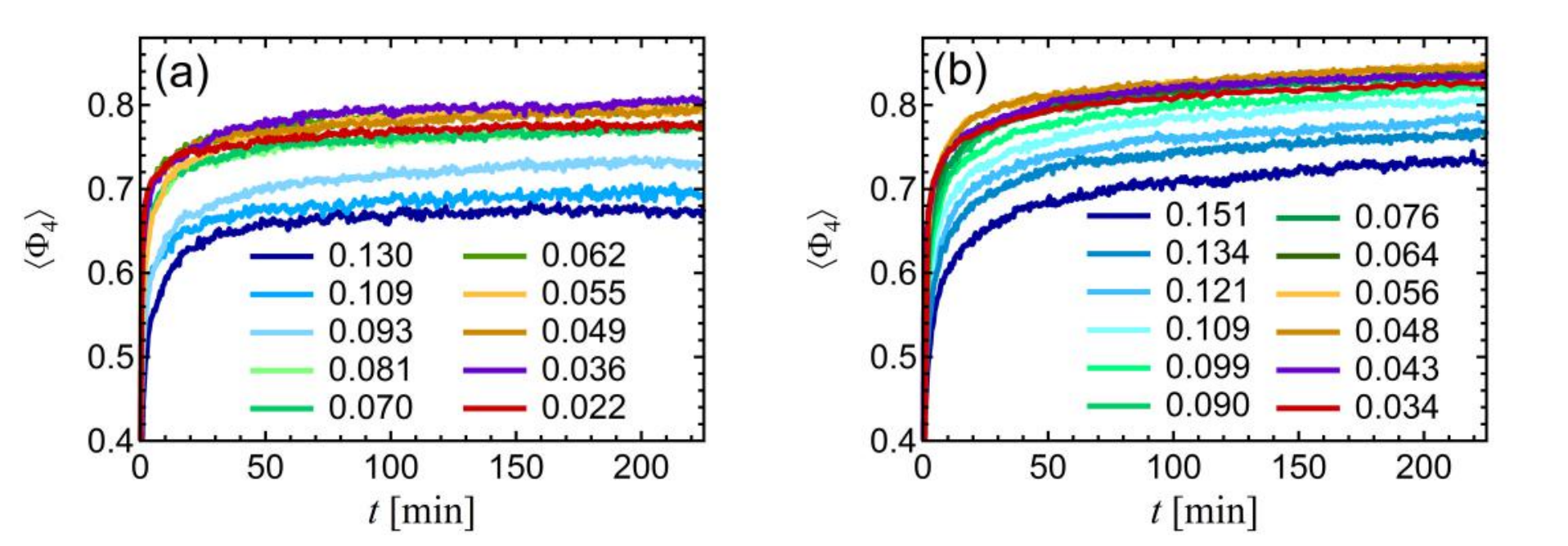}
  \caption{%
    The average bond order parameter $\langle\Phi_4\rangle$
in experiments (a) and simulations (b) are provided as a function of time for different effective temperatures.
    In each experiment, the external magnetic field was held constant for $\SI{225}{\minute}$.
    The average order parameters are calculated by taking the ensemble average of local order parameters within a given field of view as a function of time, captured at a rate of two frames per minute.   
  }
  \label{average_phi4}
\end{figure*}

\begin{figure*}[htb]
  \centering
  \includegraphics[width=0.9\textwidth]{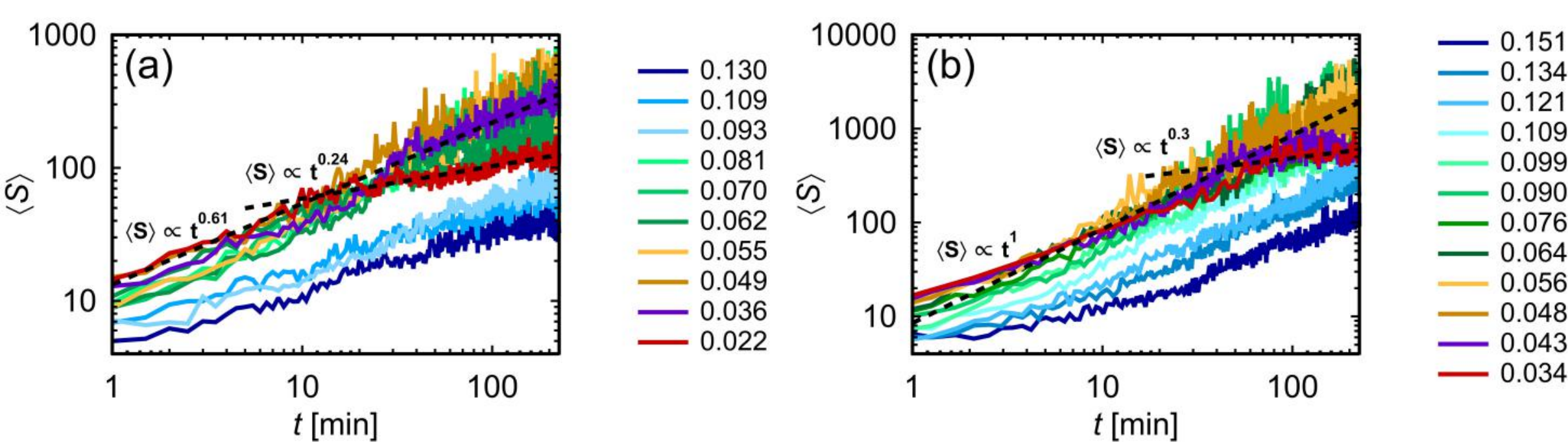}
  \caption{%
    The mean domain size $\langle S\rangle$ depicted as a function of time at different effective 
temperatures in the range $0.022 \lesssim \DLTemp \lesssim 0.130$ in experiments (a) and the range $ 0.034 \lesssim \DLTemp \lesssim 0.151$ in simulations (b).
    The system size is displayed as a log-log scale for easier visualization 
    of the power law fitting functions shown with the dashed lines.
    Domain sizes obtained in simulations are significantly larger than experiments 
    because they do not include impurities.
  }
  \label{power_law_fitting}
\end{figure*}
We use several order parameters to monitor the crystal growth process in time.
\figref{average_phi4} depicts the average bond order parameter $\langle\Phi_{4}\rangle$ as a function of time for various values of $\DLTemp$ in experiments (a) and simulations (b).
The plot shows that $\langle\Phi_{4}\rangle$ rapidly intensifies and 
reaches almost $\SI{90}{\percent}$ of the maximum value 
within the first 20 minutes in experiments, 
and within the first 30 minutes in simulations.
After the first 30 minutes, $\langle\Phi_{4}\rangle$ gradually increases for the remaining time in experiments and simulations. 
Though simulation and experiment do not match perfectly, they are reasonably consistent. 
The maximum average value of $\langle\Phi_{4}\rangle$ in experiments is $\SI{6}{\percent}$ smaller 
than values found in simulations, 
which is most likely due to the presence of particle impurities that hinder crystal growth. 
We do not continue our experiments past 300 minutes,
due to the onset of fluid evaporation and bubble formation, 
which perturb the fluid and damage the growing crystals.
%%%%%%%%%%%%%%%%%%%%%%

The crystal growth process can also be analyzed by plotting the mean domain size as a function of time on a log-log scale, 
where the linear slope indicates a power law growth process.
\figref{power_law_fitting} exhibits the time evolution of the mean domain size during rapid quenching 
for different values of the applied magnetic field $H$.
Since the domain sizes in simulations (${\sim}\,5000$ particles) tend to be significantly larger than 
in experiments (${\sim}\,600$ particles), \figref{power_law_fitting}a and \figref{power_law_fitting}b are plotted on different logarithmic scales. 
The discrepancy between experiment and simulation is thought to arise from the influence of impurities 
which are present in experiments, but not in simulations.
Similar to the results presented in \figref{average_phi4}, 
the largest crystals form in the range of $\SI{8.5}{Oe} \lesssim H \lesssim \SI{10.5}{Oe}$ ($ 0.036 \lesssim \DLTemp \lesssim 0.055$), which corresponds 
to the global maximum in the bond order parameter.
When the sample is quenched at a low temperature around $\DLTemp \approx 0.022$,
the mean domain size is limited to around ${\sim}\,100$ particles. 
These observations highlight the importance of finding an optimal condition for obtaining large single crystals.
%%%%%%%%%%%%%%%%%%%%%%%%%%%%%%%%%%%%%%%%%%%%%%%

In most simulations and experiments, we observe several distinct power laws, which are thought to be associated with different crystal growth regimes ranging from diffusion-limited to interface-limited growth kinetics.\cite{he_1997,ackerson_1995,Bohn_2010}  
These two growth regimes are clearly visible at low temperatures ($\DLTemp \lesssim 0.034$) in both experiments and simulations.  
The power law associated with the first growth regime is significantly larger than that of the second growth regime.  
To grow larger single crystals, it is important both to increase the magnitude of the power law as well as the dwell time in this first growth regime.  
At low temperatures, the crossover time between the two growth regimes typically occurs in the first 10--80 minutes. 
However, at optimal temperatures in the range of $0.036 \lesssim \DLTemp \lesssim 0.081$, we only observe one regime of growth kinetics. 
Again, we note that the best fitting power laws in experiments are always smaller than those arising in simulations due to the presence of impurities.
%%%%%%%%%%%%%%%%%%%%%%%%%%%%%%%%%%%%%%

\begin{figure*}[htb]
  \centering
  \includegraphics[width=\textwidth]{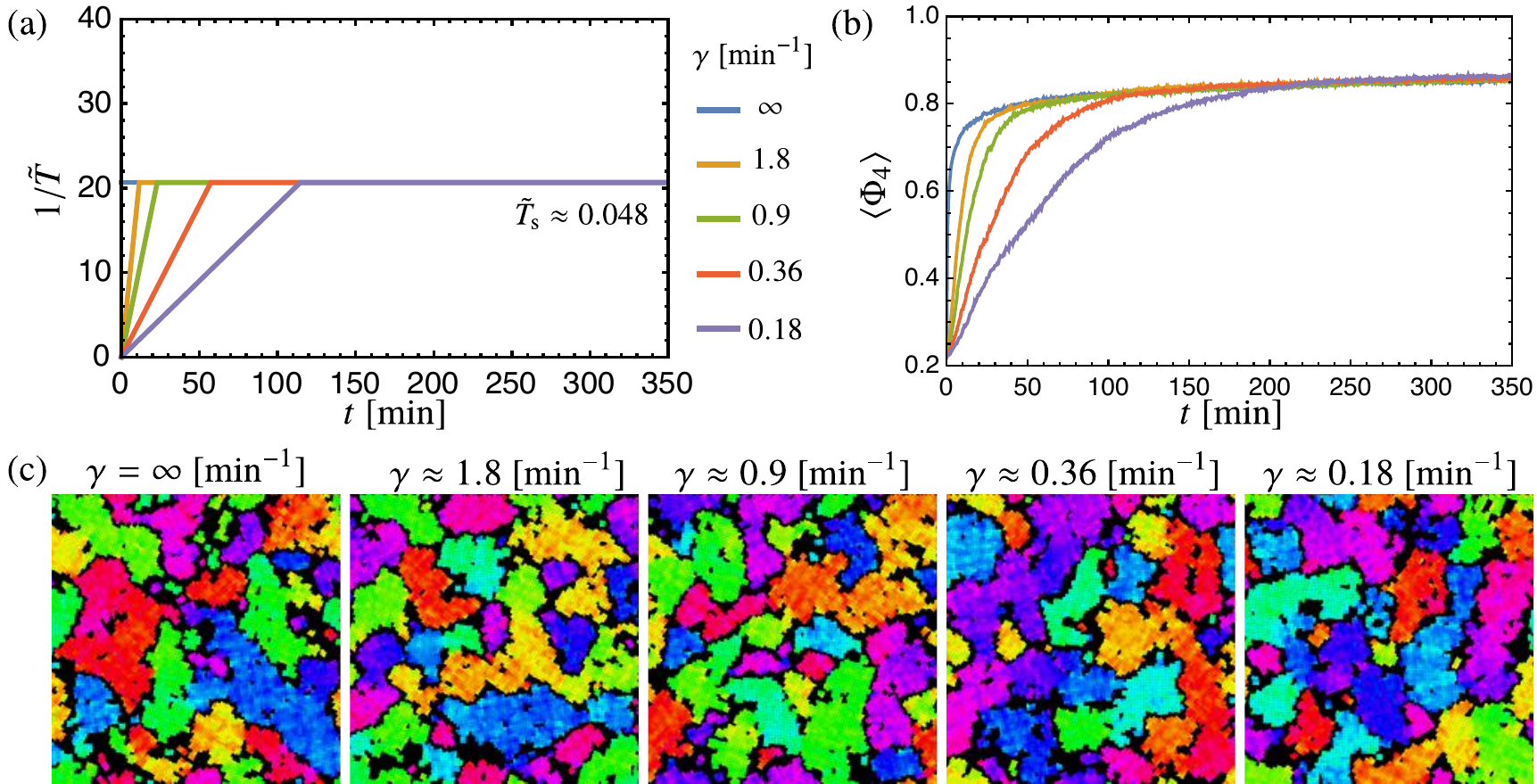}
  \caption{
  The temperature ramp rates used in simulations (a) all converge to a similar value at long times (b).  
  	Snapshots of the domain structures at the end of the simulated 350 minute time interval 
  	are shown for the different temperature ramp rates (c) .
	The polycrystals are plotted at $t=\SI{350}{\minute}$ 
	with the same scale and color scheme shown in \figref{crystal_domain_colored}.}
  \label{ramp_test}
\end{figure*}

To check the effect of the cooling rate, we conduct a series of simulations with the ramping temperature defined in accord with
\begin{equation}
  \tilde{T}(t) =
\begin{cases}
1/\gamma t,  & 0 < t <  (\gamma  \tilde{T}_{\mathrm{s}})^{-1}, \\
 \tilde{T}_{\mathrm{s}}, & 
t \geq  (\gamma  \tilde{T}_{\mathrm{s}})^{-1}, \\
\end{cases}\label{110305_18Jun16}
\end{equation}
where $\tilde{T}_{\mathrm{s}}$ is 
the setting point of effective temperature chosen to be $\DLTemp_{\mathrm{s}} \approx 0.048$, and $\gamma$ is the linear temperature ramp ranging between \SI{0.18}{\per\minute} and \SI{1.8}{\per\minute}.
Surprisingly, the average order parameters are indistinguishable from those obtained 
by rapid quenching to the same final temperature (\figref{ramp_test}b) for all values of $\gamma$.
The polycrystal domains shown in \figref{ramp_test}c also show negligible differences, which indicates that the growth kinetics are not influenced by the cooling rates.
Nonetheless, these simulations cover only the short time intervals accessible 
in experiment and are limited to only a specific temperature setting. 
More exhaustive simulations of different cooling rates and final temperatures 
may lead to even larger crystals than those observed in our experiments.

\begin{figure}[htb]
  \centering
  \includegraphics[width=0.6\columnwidth]{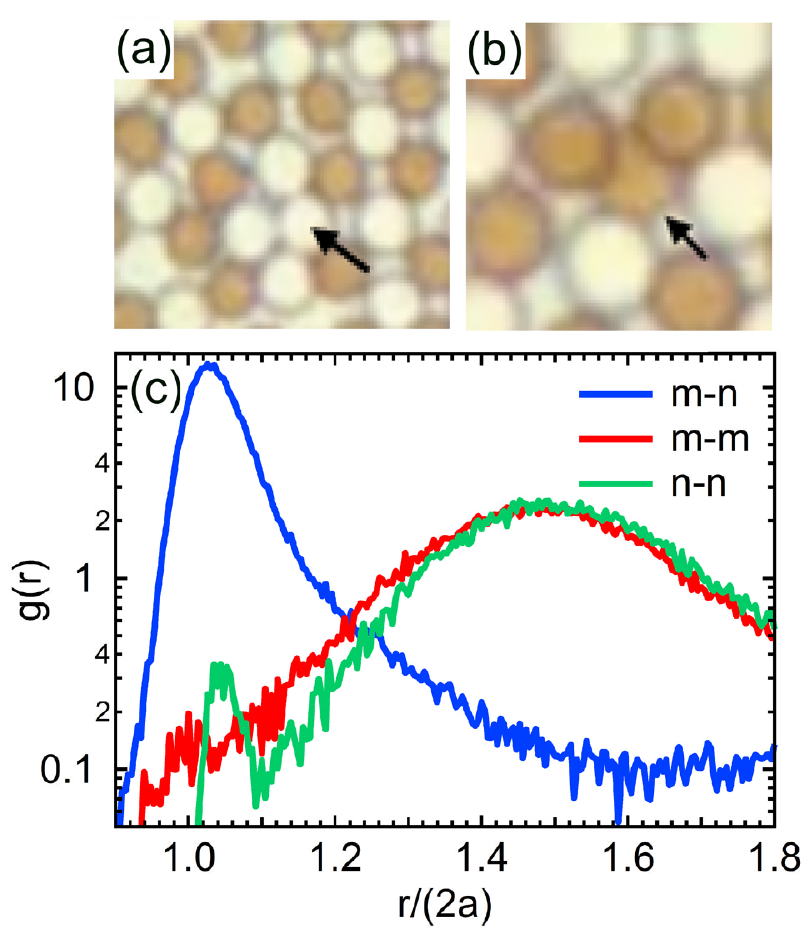}
  \caption{%
    Irreversible binding between non-magnetic particles
    (a) and the vertical stacking of particles (b), respectively. 
    The pair correlation $g(r)$ of magnetic-nonmagnetic (m-n), magnetic-magnetic (m-m), 
    nonmagnetic-nonmagnetic (n-n) are shown for a typical experiment 
    $H=\SI{7.5}{Oe}$ ($\DLTemp \approx 0.07$) 
    after $\SI{225}{\minute}$ of rapid quenching (c). 
    A log scale is applied to the vertical axis to compress the data and better visualize the peaks.
    }
  \label{impurities_pair_correlations}
\end{figure}
%%%%%%%%%%%%%%%%%%%%%%%%%%%%%%%%%%%%%%%%%%%%%%%%%%%%%%%%%%%%%%%%%%%%%%%%%

\subsection{Effect of impurities}
\label{Effect_of_defects}

%%%% figure*

In contrast to experiments, simulations allow for the consideration of systems in which all particles are identical in size, perfectly dispersed, at equal population fractions, and free of other types of interactions.
Under such conditions domains grow continuously up to several thousand particles on average, which is much larger than in experiments, where the presence of small particle clumps and occasional giant particles ($1.3$--$1.8$ times larger than the normal particles), as well as a slight imbalance between the number ratio of particles restricts the maximum domain size.   
As previously discussed, the nonmagnetic particles often have pre-existing clumps that frustrate crystal growth (\figref{impurities_pair_correlations}a). 
\figref{impurities_pair_correlations}c shows an abnormal peak in the pair correlation \eqref{115529_8Apr16} between nonmagnetic particles (n-n), which is found at a distance of $1.04$ particle diameters and is indicative of doublets formed from nonmagnetic particles within the sample.
Since direct contact between like particles is energetically unfavorable, 
we use the pair correlation function $g(r)$ as a qualitative indicator of irreversible binding between nonmagnetic particles.

\begin{figure*}[htb]
  \centering
  \includegraphics[width=0.8\textwidth]{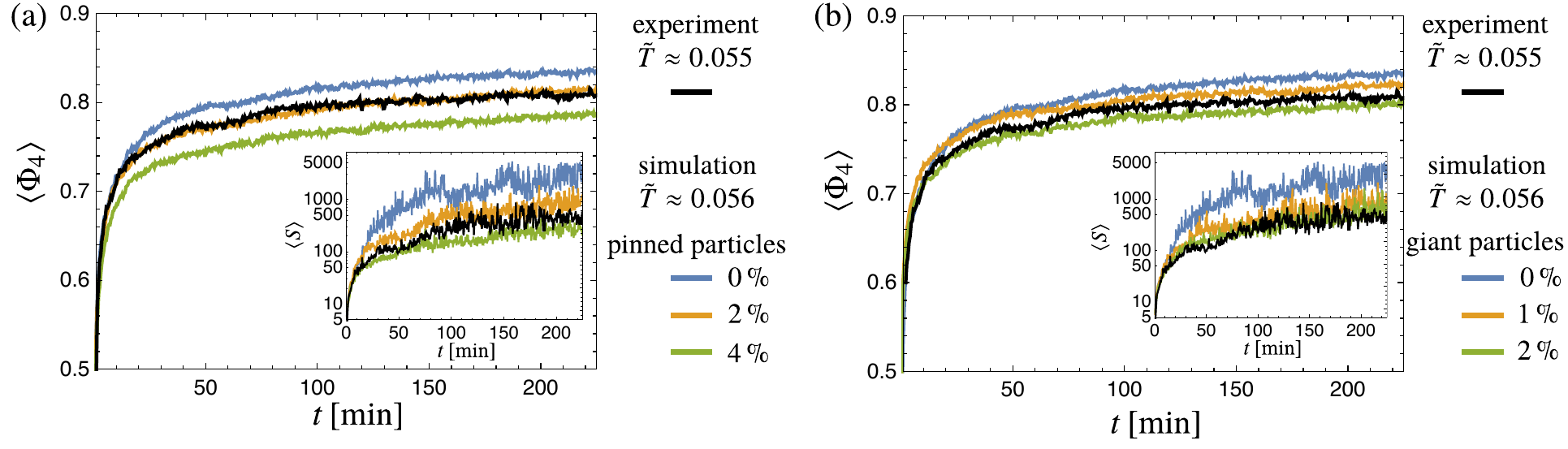}
  \caption{%
    The average bond order parameter $\langle \Phi_4 \rangle $ 
    and mean domain size $\langle S \rangle$ (insets) are shown as a function of time in simulations 
    that include up to \SI{4}{\percent} impurities of either pinned particles (a) or giant particles (b). 
    In all simulations, the particle area fraction is fixed at $\phi_{\mathrm{area}} = 0.68$ and 
    the temperature is kept at $\DLTemp \approx 0.056$.  
    The black line shows the experimental data obtained in similar conditions.  
    The best match between experiment and simulations is found when simulations 
    include approximately \SI{2}{\percent} impurities, regardless of its specific type.} 
  \label{simulation_with_defects}
\end{figure*}
%

%%%% figure*
\begin{figure*}[htb]
  \centering
  \includegraphics[width=0.8\textwidth]{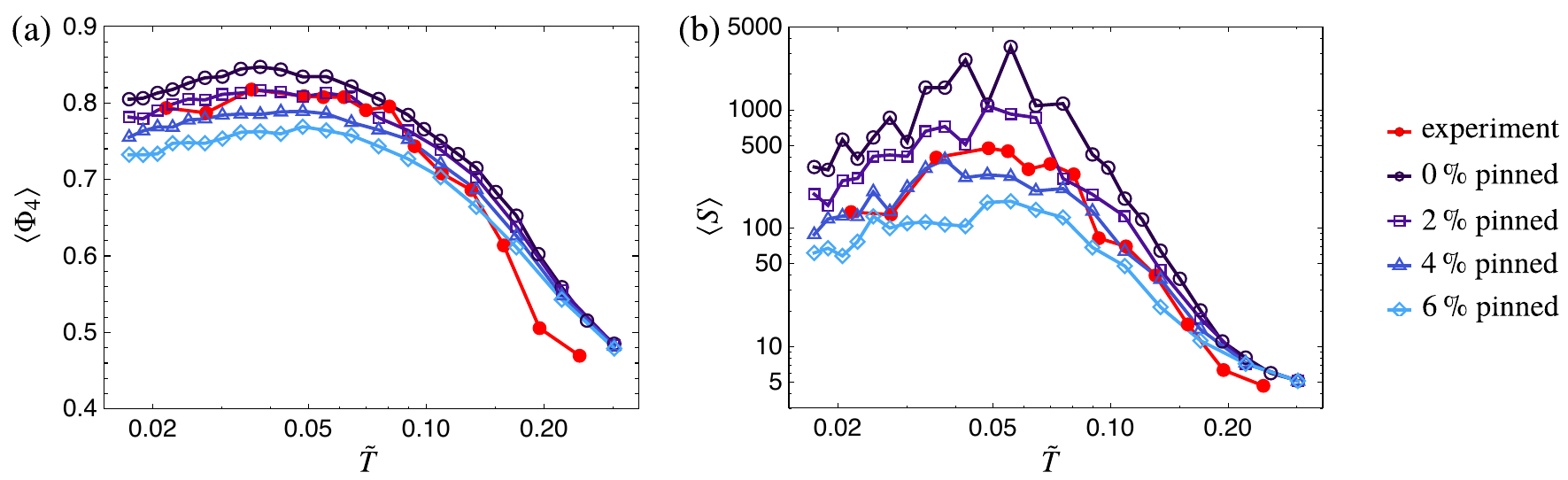}
  \caption{%
  The average bond order parameter $\langle \Phi_4 \rangle $ (a) 
  and mean domain size $\langle S \rangle$ (b) of the system are presented as a function 
  of temperature after 225 minute of elapsed time.
  Simulations include different amounts of pinned particles ranging from \SI{0}{\percent} to \SI{6}{\percent}, 
  and are compared with experimental data (red) with similar growth conditions. 
  In all simulations the particle concentrations are held at $\phi_{\mathrm{area}} = 0.68$, 
  whereas in experiments the particle concentrations were kept at $\phi_{\mathrm{area}} \approx 0.66 \pm 0.02$.  
  The impurity-free simulations are shown in black, whereas the simulations that include impurities 
  are colored from dark to light blue.  
  The best match with experiment across all temperatures is obtained for simulations that include approximately 
  \SI{2}{\percent} pinned particles.}
  \label{Phi4_S_T_dependence_exp_and_simu}
\end{figure*}

%%%%%%%%%%%%%%

Another source of discrepancy between simulation and experiment is the possibility that particles may rise out of the monolayer plane (\figref{impurities_pair_correlations}b).
Although this occurs in our experiments, it is not permitted in our simulations. 
Our image processing algorithms are not optimized to recognize stacked particles; 
thus, we do not observe any unusual correlation peak of the magnetic particles.
However, we can sometimes see this effect in experimental images 
such as those shown in \figref{impurities_pair_correlations}b. 
%

%%%%%%%%%%%%%%%%%%%%%%%%%%%%%%%%%%%%%%%%%%%%%%%%%%%%%%%%%%%%%%%%%%%%%%%%%%%%%%%%%%%%%%%%%%%%%%%%%%%%

Although we do not develop an algorithm to precisely quantify the concentration of impurities in our experimental system, 
we are able to estimate the impurity concentration to be in the range of $\SIrange{1}{3}{\percent}$ of particles, 
including ${\sim}\,\SI{0.1}{\percent}$ large particles, ${\sim}\,\SI{0.1}{\percent}$ pinned particles, 
${\sim}\,\SI{1}{\percent}$ doublets formed between nonmagnetic particles, 
and a typical number ratio of 51:49 between the magnetic and nonmagnetic particles, which vary from sample to sample.     

%%%%%%%%%%%%%%%%%%%%%%%%%%%%%%%%%%%%%%%%%%%%%%%%%%%%%%%%%%%%%%%%%%%%%%%%%%%%%%%%%%%%%%

In an attempt to better understand the effect of impurities in experiments, we include two types of impurities in our simulation: (i) particles pinned to substrates, and (ii) giant particles. 
For the sake of simplicity, we ignore particle clumps to avoid issues with trajectory analysis of rigidly linked bodies. 
As in our experiments, we perform simulations in which the particle concentration is kept constant at $\phi_{\mathrm{area}} = 0.68$, 
while the percentage of impurities is varied in the range of $\SIrange{0}{6}{\percent}$ 
and is assumed to be initially randomly distributed in the simulation box.
Giant particles are assumed to have diameters 1.5 times that of other particles.
On the other hand, pinned particles are assumed to be the same diameter as other particles, but are spatially fixed.
The impact of impurities on the average bond order parameter and the mean domain size in the case 
of $\DLTemp \approx 0.055$ are shown in \figref{simulation_with_defects}.
As expected, these quantities both decay significantly as the number of impurities is increased.
The best match with experiment is achieved when simulations employs a impurity concentration of $\SI{2}{\percent}$, which is consistent with the level observed in experiments. 

%%%%%%%%%%%%%%%%%%%%%%%%%%%%%%%%%%%%%%%%%%%%%%%%%%%%%%%%%%%%%%%%%%%%%%%%%%%%%%%%%%%%%%%%%%%%%%%%%%%%

\figref{Phi4_S_T_dependence_exp_and_simu} shows 
the average bond order parameter $\langle \Phi_{4} \rangle$ and 
the mean domain size $\langle S \rangle $ 
as a function of the quenching temperature after 225 minutes of elapsed time. 
Simulation results are obtained by varying the concentrations of pinned particles 
in the range of $\SIrange{0}{6}{\percent}$.
Similar to what is shown in \figref{simulation_with_defects}, the best match with experimental results 
is obtained when simulations include an impurity concentration of approximately \SI{2}{\percent}.
Near the effective solidification temperature, the slight mismatch between experiment and simulation is not unexpected due to the influence of unknown short-range particle interactions (namely surface charge repulsive forces and other surface forces), which are not included in our simulations. 
These interactions should be negligible when the magnitude $H$ of the applied magnetic field is sufficiently large, where an improved quantitative agreement between experiments and simulations is observed. 
%

%%%%%%%%%%%%%%%%%%%%%%%%%%%%%%%%%%%%%%%%%%%%%%%%%%%%%%%%%%%%%%%%%%%%%%%%%%%%%%%%%%%%%%%%%%%%%%%%%%%%

Matching the experimental curve to simulations leads to a best fitting parameter for the magnetic permeability of the ferrofluid  determined from \eqref{particle susceptibility} with $ \bar\chi_{\mathrm{n}} \approx -0.36$, which yields $\mu_{\mathrm{f}} \approx 1.47 \mu_{\mathrm{0}}$.  
According to \eqref{ferrofluid permeability}, this best fitting estimate of the ferrofluid permeability implies that the bulk susceptibility of 
the material constituting the nanoparticles is $ \chi_{\mathrm{b}} \approx 47$.  
%nanoparticle material
%
The permeability of the magnetic bead can be estimated similarly from \eqref{ferrofluid permeability} with $\bar\chi_{\mathrm{m}} \approx 0.36$, which yields $\mu_{\mathrm{m}} \approx 2.07\mu_{\mathrm{0}}$.  
This value suggests that the magnetic susceptibility of the magnetic particles in vacuum is $\bar\chi_{\mathrm{m}} \approx 1.07$, which is larger than the typically reported values of these bead types in water falling in the range of $0.17$--$0.96$.\cite{Tierno_2007,Du_2013,Tierno_2016}  
To facilitate comparisons with our prior paper,\cite{Yang_2015} in which a correction factor $\alpha=2.4 \pm 0.3$ was used, that acted as multiplier in the particle susceptibilities, we estimate the best fitting parameters for the magnetic permeabilities of the magnetic particles and ferrofluid to be $\mu_{\mathrm{m}} \approx 2.56\mu_{\mathrm{0}}$ and $\mu_{\mathrm{f}} \approx 1.66\mu_{\mathrm{0}}$, respectively.
Thus, the magnetic parameters of the present study provide a slightly better approximation of the expected values in vacuum.
%

%%%% figure*
\begin{figure}[htb]
  \centering
  \includegraphics[width=0.5\textwidth]{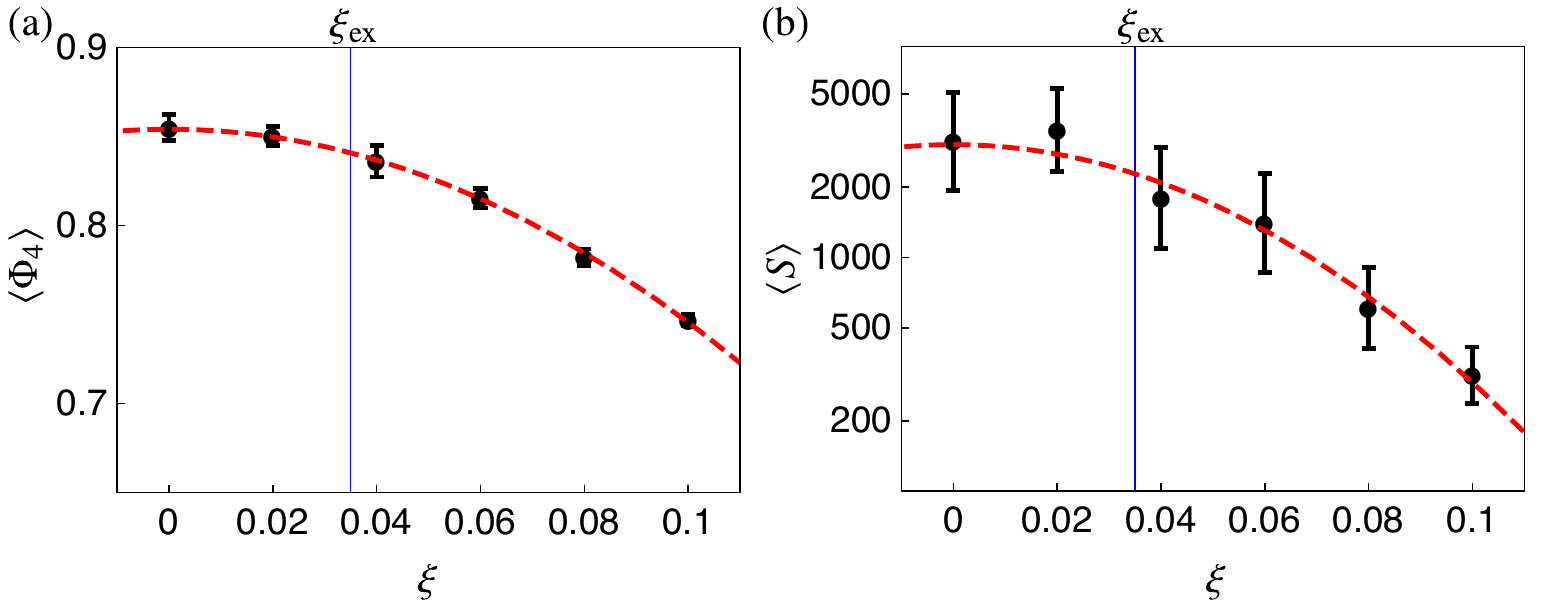}
  \caption{%
    The average bond order parameter $\langle \Phi_4 \rangle $, 
    and mean domain size $\langle S \rangle$
    are plotted as a function of the relative particle concentration ratio in the range of $\xi = \numrange{0}{0.1}$.  
    The simulations were averaged over ten independent runs 
    in order to obtain the mean and standard deviations. 
    All simulations were conducted for 225 minutes and at 
    a temperature of $\tilde{T} \approx 0.048$. 		
    The vertical blue lines show the average population imbalance of the typical experiments: $\xi_{\mathrm{ex}} \approx 0.035 $.
    The red dashed curves are provided as a guide to the eye.
  } 
  \label{nr_dependence}
\end{figure}
%

%%%% figure*
\begin{figure}[htb]
  \centering
  \includegraphics[width=\columnwidth]{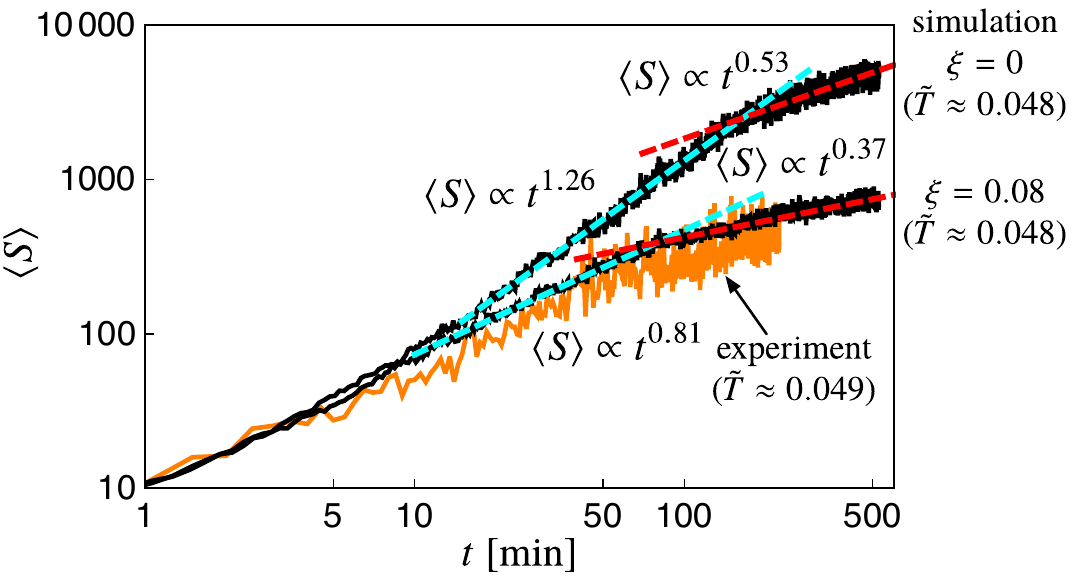}
  \caption{%
    The two regimes of growth kinetics are shown 
    for a perfectly balanced system ($\xi = 0$) 
    and for a system that contains an excess of one particle type corresponding to $\xi = 0.08$, 
    with the effective temperature fixed at $\DLTemp \approx 0.048$.  
    Simulations (black lines) were averaged for ten independent runs, 
    whereas the result from one experiment with growth conditions 
    similar to simulations is shown in orange. 
    The best fitting power laws 
    for the first and second growth regimes are shown with the blue dotted 
    and red dotted lines, respectively. 
  }
  \label{time_evolution_xi0_xi008_and_experiment}
\end{figure}

Despite our best effort to balance the relative particle concentrations,
the numbers $N_{\mathrm{m}}$ and $N_{\mathrm{n}}$ of magnetic and nonmagnetic particles differ in experiments.
To examine the influence of this impurity type, we simulate particle concentration differences $\xi$ ranging from $0$ to $0.1$,
where $\xi \equiv (N_{\mathrm{m}} - N_{\mathrm{n}})/(N_{\mathrm{m}} + N_{\mathrm{n}})$.
The effects of the population imbalance 
on the average bond order parameter $\langle \Phi_{4} \rangle$ (\figref{nr_dependence}a) 
and the mean domain size $\langle S \rangle$ (\figref{nr_dependence}b) are depicted after a 225 minute elapsed time interval 
in a constant temperature of $\DLTemp \approx 0.048$.
As expected, both $\langle\Phi_4\rangle$ and $\langle S\rangle$
are observed to decrease with increasing population imbalance. 
This effect is not surprising, because an excess of one particle type leads to a non-zero net dipole moment of the crystal, which induces stress and hinders crystal growth.\cite{yang2013tunable}
At the ratios typically observed in experiments, $\xi_{\mathrm{ex}} = 0.035 \pm 0.018$, 
the values of $\langle\Phi_4\rangle$ and $\langle S\rangle$ 
are slightly smaller than those obtained under ideal simulation conditions ($\xi = 0$).
This value of $\xi$ corresponds to one particle type being present 
at an excess of approximately \SI{3}{\percent}, which is in a similar range to the other types of considered impurities. 

%%%%%%%%%%%%%%%%%%%%%

The influence of population imbalance on the growth kinetics is shown in \figref{time_evolution_xi0_xi008_and_experiment}, 
which depicts $\langle S \rangle$ as a function of time.  
Two crystal growth regimes can be distinguished in both the perfectly balanced system and in the system containing an excess of one particle type with $\xi =0.08$. 
Comparing these corresponding growth curves, we find several features demonstrating how the perfectly balanced system attains larger crystalline domain sizes than the system containing a particle imbalance.  
First, in the perfectly balanced system the power law of the first growth regime attains a higher value ($1.26$) than that found in the system including an excess ($0.81$) of one particle type.  
This more rapid initial growth process allows the perfectly balanced system to reach larger crystal sizes before entering into the ripening phase of the second growth regime.  
Second, in the perfectly balanced system the cross-over time between the different growth regimes is delayed in comparison to what occurs for the system containing a particle imbalance.  
Thus, the perfectly balanced system has more time to grow domains prior to entering the second regime with slower growth kinetics. 
For comparison, we also provide one experimental plot for a system with similar growth conditions as the simulation with an assumed population balance $\xi =0.08$. 
The data from this single experiment contains more noise than the data arising from simulations, which is averaged over 10 individual trials.  
%%%%%%%%%%%%%%%%%%%%%%%%%%%
 
Based on our simulations, we conclude that the number rather than the specific type of impurities is the determining factor for matching with experiment. 
Thus, instead of developing a complex model, which incorporates multiple kinds of impurities in one simulation, we instead focus on a single impurity type. 
In \figref{simulation_with_defects} and \ref{Phi4_S_T_dependence_exp_and_simu}, we show that including a small percentage of pinned or giant particles can provide a match with experiments. 
In \figref{time_evolution_xi0_xi008_and_experiment}, we also show that an imbalance in the particle number ratio can also match experiments at the effective temperature ($\tilde{T} \approx 0.049$) and timescale ($t = \SI{225}{\minute}$), with an assumed population imbalance $\xi = 0.08$.  
It is thus clear that a small percentage of defects can dramatically influence the domain sizes of 2D crystals.
%
%#!latexmk main.tex
\section{Conclusions}

We developed an experimental system for forming two-dimensional,
close-packed, colloidal crystal alloys with more than ${\sim}\,1000$ particles 
in a single domain. 
Potentially optimal annealing kinetics were achieved by simultaneously controlling 
the density and particle interactions within a 2D monolayer of magnetic 
and nonmagnetic particles immersed in a ferrofluid.
Specifically, the best crystals were obtained for area fractions of $\SIrange{65}{70}{\percent}$, 
which are small enough to avoid glass or jamming transitions occurring near 
the ideal packing fraction of a square lattice (${\sim}\SI{78}{\percent}$) 
while being large enough to prevent voids, which are common below an area fraction of $\SI{60}{\percent}$.
We also found that the largest crystals were obtained for magnetic fields with strengths in the range of $\SI{8.5}{Oe} \lesssim H \lesssim \SI{10.5}{Oe}$, 
which are high enough to overcome thermal fluctuations while being low enough to allow efficient annealing of domain boundaries and to remove impurities from crystal interiors.

%%%%%%%%%%%%%%%%%%%%%%%%%%%%%%%%%%%%%%%%%%%%%%%%%%%%%%%%%%%%%%%%%%%%%%%%%%%%%%%%%%%%%%%%%%%%%%%%%%%%%%%%%%%%%%%%%%%%%%%%%%%%%%%%%%%%%%%%%%%%%

Brownian dynamics simulations based on a point dipole model were found able to reproduce the main features of the assembly process; however, they predict much larger crystals than are obtained in experiments.
Results obtained from simulations of an idealized, impurity-free system agreed qualitatively but deviate quantitatively from experiment, where a small percentage of particles that form clumps and adhere to the substrate are generally present.
An imbalance in the particle number ratio also led to smaller crystals, mimicking the effect of the other experimental impurities. 
We found that including a similar percentage of impurities in simulations ($\SIrange{1}{2}{\percent}$) yields better quantitative agreement between 
theory and experiment. 
Additionally, we achieved a more precise characterization of the magnetic susceptibility of the magnetic bead $\bar{\chi}_{\mathrm{m}} = 1.07$, 
which is $\SI{30}{\percent}$ smaller than in our prior analyses.\cite{Yang_2015}

%%%%%%%%%%%%%%%%%%%%%%%%%%%%%%%%%%%%%%%%%%%%%%%%%%%%%%%%%%%%%%%%%%%%%%%%%%%%%%%%%%%%%%%%%%%%%%%%%%%%%%%%%%%%%%%%%%%%%%%%%%%%%%%%%%%%%%%%%%%%

While the mean domain size we were able to experimentally obtain is on the order of ${\sim}\,600$ particles, 
future improvements of the experimental conditions through the use of better passivated surfaces, the reduction of particle aggregation and imbalance in particle concentration should enable us to approach crystal sizes of ${\sim}\,5000$ particles predicted in the idealized simulations. 
The ability to obtain sufficiently large colloidal crystal alloys 
will facilitate future scientific investigations on grain boundary motions,\cite{Edwards_2014} dynamics of vacancies and defects,\cite{Pertsinidis_2001} jamming transitions,\cite{Stratford_2005} and various types of phase transitions that are difficult or impossible to observe in atomic systems.

%#!latexmk main.tex
%%%1. Introduction%%%%%%%%%%%%
\section*{Acknowledgement}
The authors are thankful for support from the National Science
Foundation Research Triangle Materials Research Science and
Engineering Center (DMR-1121107) and the Okinawa Institute of Science and Technology Graduate University with subsidy funding from the Cabinet Office, Government of Japan.
%%%%%%%%%%%%%%%%%%%%%%%%%%%%%%%%%%%%%%%%%%%%%%%%%%%%%%%
%\appendix
%\input{a1_hydrodynamic_interaction}

%\input{main.bbl}
\bibliography{main} %your .bib file
\bibliographystyle{rsc} %the RSC's .bst file

%%%%%%%%%%%%%%%%%%%%%%%%%%
% \twocolumn
% \begin{figure}[h]
% \centering
%   \includegraphics[height=6cm]{Phi4}
%   \caption{Order parameter as a fucntion of field strength}
%   \label{fgr:Phi4}
% \end{figure}
\end{document}